\documentclass[aps,pra,twocolumn,showpacs,longbibliography]{revtex4-2}
\usepackage[plainpages=false,pdfpagelabels,colorlinks=true,linkcolor=red,urlcolor=blue,citecolor=blue,pdftitle={Title},pdfauthor={},pdfdisplaydoctitle=true,pdfduplex=DuplexFlipLongEdge]{hyperref}
\usepackage{amsmath, amsthm, amssymb}
\usepackage{array}
\usepackage{bbold} 
\usepackage{bm}        
\usepackage[dvipsnames,usenames]{color}
\usepackage{dcolumn}
\usepackage{float}
\usepackage{graphicx}  
\usepackage{lipsum}
\usepackage{mathrsfs}
\usepackage{mathtools,amssymb,graphicx,units,bm}
\usepackage{physics}%
\usepackage{soul} 
\usepackage{booktabs}
\usepackage[normalem]{ulem}

\graphicspath{{figures/}}

\emergencystretch=\maxdimen
\usepackage[utf8]{inputenc}
\DeclareUnicodeCharacter{202F}{\,} 

\begin{document}

\normalem

\title{\large \textbf{Could the 650 GeV Excess be a Pseudoscalar of a 3-Higgs Doublet Model?}}
\author{Ayoub Hmissou}
\affiliation{Laboratory of Theoretical and High Energy Physics,Faculty of Science, Ibnou Zohr University, B.P 8106, Agadir, Morocco.}
\author{Stefano Moretti}
\affiliation{School of Physics \& Astronomy, University of Southampton, Southampton, SO17 1BJ, United Kingdom}
\affiliation{Department of Physics \& Astronomy, Uppsala University, Box 516, SE-751 20 Uppsala, Sweden.}
\author{Larbi Rahili}
\affiliation{Laboratory of Theoretical and High Energy Physics, Faculty of Science, Ibnou Zohr University, B.P. 8106, Agadir, Morocco.}

\date{Version 5.0 -- \today}

\begin{abstract}
In this study, we propose the interpretation of a 650 GeV excess observed at the Large Hadron Collider (LHC) by the CMS Collaboration in terms of the production of a CP-odd (or pseudoscalar) Higgs boson $A$, with mass around 650 GeV, decaying into the Standard Model (SM)-like Higgs state $h_{125}$ (in turn decaying into $\gamma\gamma$) and a $Z$ boson (in turn decaying into $b\bar b$), within a 3-Higgs Doublet Model (3HDM) featuring two active and one inert doublet, known as the I(1+2)HDM. This theoretical structure features a spectrum with both the SM-like Higgs boson  (with a 125 GeV mass) and a lighter CP-even (or scalar) Higgs state with mass around 95 GeV, $h_{95}$, which is present in this scenario for the purpose of simultaneously explaining anomalies seen in the $b\bar b$, $\gamma\gamma$ and $\tau^+\tau^-$ final states in searches for additional light Higgs states at the Large Electron-Positron (LEP) collider and  LHC itself.  It should be noted that, in the I(1+2)HDM, the inert sector presents loop-induced enhancements to the $h_{95} \to \gamma\gamma$ width via additional scalar charged states, providing a viable mechanism to explain, in particular, the observed (and most significant) di-photon excess at 95 GeV. Taking into account both experimental and theoretical constraints, our results can not only explain the aforementioned anomalies (possibly, aside from the $\tau^+\tau^-$, which is the most marginal one) but also predict, as collateral signals, resonant production of the same CP-odd scalar $A$ followed by the decays:   
(i)  $A \to h_{95} \, Z$, leading to the same $\gamma\gamma b \bar{b}$ final state displaying the original 650 GeV anomaly and (ii) $A\to t\bar t$, leading to a well-known and studied signature. Both of these signals are potentially explorable at Run 3 of the LHC and most possibly so at the High-Luminosity LHC (HL-LHC),  while being consistent with current data at a significance level of $2.5 \sigma$.
\end{abstract}
\maketitle

\section{Introduction}
The discovery of a Higgs boson with a mass of 125 GeV by the ATLAS \cite{ATLAS:2012yve} and CMS \cite{CMS:2012qbp} Collaborations marked a triumph for the Standard Model (SM) of particle physics and has opened the door to precision Higgs physics, while at the same time motivating searches for additional (psudo)scalar states Beyond the SM (BSM). Intriguingly, several experimental hints of BSM physics have emerged in recent years, notably, a local excess near 95 GeV in the $b\bar{b}$ channel at the Large Electron-Positron (LEP) collider as extracted by all Collaborations therein (ALEPH, DELPHI, L3, and OPAL)  \cite{LEPWorkingGroupforHiggsbosonsearches:2003ing,ALEPH:2006tnd,Cao:2016uwt} plus in the $\gamma\gamma$ and $\tau^+\tau^-$ decay modes at the LHC as seen by CMS \cite{CMS:2024yhz,CMS:2022goy} and ATLAS \cite{ATLAS:2018xad}\footnote{Note that the best-fit mass  actually differs from one channel to another, however, given the poor  resolution in invariant mass of $b\bar b$ and $\tau^+\tau^-$ pairs, compared with to the $\gamma\gamma$ case (which is centered at 95 or so GeV), we collectively use the latter value throughout, including in our forthcoming tests of the various anomalies.}. Recently, a CMS search has further reported an excess in events featuring a di-photon and a bottom quark pair (i.e., a $\gamma\gamma b\bar b $ final state), which could indicate a heavier resonance $X$ with a mass of approximately 650 GeV decaying into a secondary object  $Y$ with a mass in the range of 90--100 GeV and the SM-like Higgs boson. Such an anomaly was observed with a local excess of 3.8$\sigma$ and the best-fit value given for the cross section of this excess ${\gamma \gamma b\bar{b}}$ final state was as follows \cite{CMS:2023boe}:
\begin{equation}
\sigma (pp \rightarrow X_{650} \rightarrow h_{125} Y \rightarrow \gamma \gamma b\bar{b}) = 0.333_{-0.13}^{+0.17} \,\, (\text{fb}).
\end{equation}
Such anomalies, although not yet conclusive in terms of significance individually, taken together, actually stand out as compelling candidates for new phenomena, and their collective persistence motivates the search for theoretical frameworks that can simultaneously explain them all.

Extensive theoretical work has been devoted to explaining the observed excesses around 95 GeV. These include non-supersymmetric models such as the 2-Higgs Doublet Model (2HDM) \cite{Cacciapaglia:2016tlr,Benbrik:2022azi, Benbrik:2022tlg,Khanna:2024bah}, its singlet (Next-to-2HDM (N2HDM)) and triplet extensions \cite{Biekotter:2019kde,Heinemeyer:2021msz,Biekotter:2022jyr,Banik:2023ecr,Biekotter:2023jld,Biekotter:2023oen,Kundu:2024sip}, the Georgi-Machacek (GM) model \cite{Ahriche:2023wkj,Chen:2023bqr,Du:2025eop}, the minimal gauged two-Higgs-doublet model (G2HDM) \cite{Arhrib:2024zsw}, the 3HDM containing one inert doublet (or I(1+2)HDM) \cite{Hmissou:2025uep} as well as supersymmetric frameworks such as the Next-to-Minimal Supersymmetric SM (NMSSM) \cite{Cao:2016uwt,Cao:2019ofo,Biekotter:2021qbc,Li:2023kbf,Ellwanger:2024vvs,Lian:2024smg,Ellwanger:2024txc,Cao:2024axg,Hammad:2025wst} and the so-called  $\mu\nu$SSM \cite{Cao:2023gkc}. Meanwhile,  the anomaly observed around 650 GeV has also attracted attention and has been investigated within some BSM scenarios, including the NMSSM \cite{Ellwanger:2023zjc} and 2HDM  \cite{Benbrik:2025hol}. 

In this paper, we examine yet another possibility to explain the 650 GeV anomaly, by using as a theoretical scenario the one of \cite{Hmissou:2025uep}: i.e., the I(1+2)HDM.  In fact, in doing so, we seek simultaneously a solution to the 95 GeV anomalies. As seen in \cite{Hmissou:2025uep}, the advantage of this BSM framework is that it provides an additional pair of charged scalar states, $\chi^\pm$, which enhances the one-loop decay into two photons of the lightest CP-even Higgs state, $h_{95}$, thereby relieving the tension existing in the 2HDM Type-I when attempting to maximise simultaneously the $b\bar b$, $\gamma\gamma$ and $\tau^+\tau^-$ decay rates \cite{Khanna:2024bah}. In fact, the I(1+2)HDM that we use here is of Type-I, just like in Ref.~\cite{Hmissou:2025uep}. Herein, we assume that the $X_{650}$ is the pseudoscalar Higgs state, \( A \), with a mass of 650 GeV, decaying into the SM-like Higgs boson \( h_{\rm 125} \) and a \( Z \) boson, followed by the subsequent decays $h_{\rm 125} \rightarrow \gamma\gamma$ and  $Z \rightarrow b\bar{b}$, respectively. In particular, we are suggesting that the observed $b\bar{b}$ pair originates from a \( Z \) boson decay, which would then take the place of the corresponding decay of the lightest Higgs boson with a mass around 95 GeV (the one used in the NMSSM explanation), indeed, like in Ref.~ \cite{Benbrik:2025hol}. 

Notice that this interpretation is well motivated. In fact, although the CMS analysis~\cite{CMS:2023boe} originally models the excess as a decay $X_{650} \to h_{125} Y_{95}$, with $Y_{95}$ being a spin-0 state, at $\sqrt s =m_A \simeq 650$~GeV, where $m_Z^2/s \ll 1$, the equivalence theorem ensures that the psudoscalar polarisation of the $Z$ boson behaves like the corresponding neutral Goldstone mode, with deviations suppressed by ${\cal O}(m_Z^2/s) \lesssim 2\%$. Hence, also considering the limited mass resolution of $b\bar b$ pairs, of some 10 GeV or so, and the fact that none of the selection cuts used by CMS has a marked spin dependence, it is conceivable that the $Z$ boson of the SM is behind the 95 GeV component of the 650 GeV anomaly. 

Additionally, while assuming this excess, one should bear in mind that the $\tau^+\tau^-b\bar{b}$ channel serves as a valuable complementary probe of the same underlying dynamics. Notably, CMS has performed a dedicated search for heavy resonances in this final state, setting a $95\%$ Confidence Limit (CL) upper limit of approximately 3 fb on the cross section $\sigma (pp \rightarrow X_{650} \rightarrow \tau^+\tau^- b\bar{b})$ \cite{CMS:2021yci}, which would then constrain {the $pp \rightarrow A  \rightarrow h_{\rm 125} Z ({\rm or}~Z h_{\rm 125}Z\to  \tau^+\tau^- b\bar{b}$ processes in our I(1+2)HDM.} We have thus verified that the parameter space of the I(1+2)HDM, providing the illustrated explanation to all discussed anomalies, is also compliant with the aforementioned limit. The generated parameter space points were further tested against the most recent CMS upper limits on the cross section for the process $pp\to Y(\to\tau^+\tau^-)H(\to\gamma\gamma)$
\cite{CMS:2025tqi} and these were found to constrain significantly the I(1+2)HDM parameter space (see below). Further bearing in mind \cite{CMS:2025qit} alongside \cite{CMS:2025tqi}, the following production and decay patterns were also subject to scrutiny: $pp\to h_{125}(\to b\bar b)Y(\to\gamma\gamma)$ and $pp\to h_{125}(\to\tau^+\tau^-)Y(\to\gamma\gamma)$. However, given that for our solution we have $Y\equiv Z$~\footnote{And the I(1+2)HDM used here is CP-conserving, so that the decay $A\to h_{125}h_{95}$ is not possible.}, the (non-resonant) {\sl transition} $Z^*\to\gamma\gamma $  is highly suppressed because of the Landau-Yang theorem
\cite{Landau:1948kw,Yang:1950rg} (note also Ref.~\cite{Moretti:2014rka}),  the corresponding experimental limits quoted by CMS are irrelevant to our theoretical scenario (so we ignore these in the remainder).

Finally, in order to test our theoretical hypothesis, we also make predictions for two additional processes that would emerge over the same parameter space of the I(1+2)HDM, explaining all the aforementioned anomalies, both of which may be testable at the present and/or future stages of the LHC. These are the channels $pp\to A \to h_{95} \, Z$ and $pp\to A\to t\bar t$.

The layout of the paper is as follows.  In Section \ref{sec:setup-constraints}, we briefly review the main features of the I(1+2)HDM, along with the considered experimental and theoretical constraints. 
In Section \ref{sec:res-disc} we present our main results by illustrating the correlations among Higgs boson masses and production cross sections relevant to searches for additional heavy resonances in various final states. In the last section, we summarise our findings.

\section{Model Setup and Constraints}
\label{sec:setup-constraints}
In this section, we outline the I(1+2)HDM model, which includes a scalar Dark Matter (DM) candidate. We begin with a recap of the scalar potential and the theoretical and experimental constraints affecting it. Next, we provide an analytical explanation for the $\gamma\gamma b \bar{b}$ excess within this model framework.

\subsection{I(1+2)HDM Basics}
\label{sec:modelsetup}
The I(1+2)HDM model is composed of two active Higgs doublets and one inert doublet. The extension of the SM  adding only one  extra Higgs doublet has been widely studied, with the 2-Higgs-doublet model (2HDM) being one of the most explored frameworks for such an extension \cite{Branco:2011iw}. The I(1+2)HDM discussed  in Refs.~\cite{Grzadkowski:2009bt,Moretti:2015cwa} and adopted here features a discrete $\mathbb{Z}_2 \times \mathbb{Z}^\prime_2$ symmetry, where the $\mathbb{Z}_2$ symmetry  is enforced upon the inert doublet, meaning that only the inert field $\eta$ transforms as $\eta \rightarrow -\eta$, while the Higgs doublets $\Phi_1$ and $\Phi_2$ remain unaffected. In addition to this, a softly broken $\mathbb{Z}^\prime_2$  symmetry is implemented for the active Higgs fields, following which $\Phi_1$ remains unchanged whereas $\Phi_2$ undergoes the transformation $\Phi_2 \rightarrow -\Phi_2$. This specific symmetry breaking, as suggested by the Paschos-Glashow-Weinberg theorem \cite{Glashow:1976nt}, is crucial for suppressing Flavour Changing Neutral Currents (FCNCs) at tree level, ensuring the consistency of the model.

The scalar potential that remains invariant under both  the gauge group of the SM and  the two discrete symmetries introduced above is given as follows: 
\begin{widetext}
\begin{align} 
V &= -\frac12\left\{m_{11}^2\Phi_1^\dagger\Phi_1 
+ m_{22}^2\Phi_2^\dagger\Phi_2 + \left[m_{12}^2 \Phi_1^\dagger \Phi_2 
+  \text{h.c.} \right] \right\} + \frac{\lambda_1}{2}(\Phi_1^\dagger\Phi_1)^2 
+ \frac{\lambda_2}{2}(\Phi_2^\dagger\Phi_2)^2 \nonumber \\
&+ \lambda_3(\Phi_1^\dagger\Phi_1)(\Phi_2^\dagger\Phi_2) + \lambda_4(\Phi_1^\dagger\Phi_2)(\Phi_2^\dagger\Phi_1)  + \frac12\left[\lambda_5(\Phi_1^\dagger\Phi_2)^2 + \text{h.c.} \right]\nonumber \\
&+ m_\eta^2\eta^\dagger \eta + \frac{\lambda_\eta}{2} 
(\eta^\dagger \eta)^2 + \lambda_{1133} (\Phi_1^\dagger\Phi_1)(\eta^\dagger \eta)
+\lambda_{2233} (\Phi_2^\dagger\Phi_2)(\eta^\dagger \eta)\nonumber \\
& +\lambda_{1331}(\Phi_1^\dagger\eta)(\eta^\dagger\Phi_1) 
+\lambda_{2332}(\Phi_2^\dagger\eta)(\eta^\dagger\Phi_2) 
+\frac{1}{2}\left[\lambda_{1313}(\Phi_1^\dagger\eta)^2+\lambda_{2323}(\Phi_2^\dagger\eta)^2 +\text{h.c.} \right]
\label{pot} 
\end{align}
\end{widetext}
where the $\lambda_i$'s denote the quartic coupling  parameters while $m_\eta^2$, $m_{11}^2$, $m_{22}^2$ and $m_{12}^2$ are the mass-squared terms.

The two active Higgs doublets and the inert doublet under $SU(2)_{L}$  are parameterised as follows, respectively:
\begin{equation} 
\Phi_k=\left(
\begin{array}{c}
\phi_k^\pm   \\ 
(v_k+\eta_k+i z_k)/\sqrt{2}
\end{array}\right),  \quad
\eta = \left(
\begin{array}{c}
 \chi^\pm \\ 
 (\chi + i \chi_a)/\sqrt{2} 
\end{array}
\right).
\label{doublets}
\end{equation}

Within this framework, $v_1$ and $v_2$ denote the non-zero Vacuum Expectation Values (VEVs) developed by the neutral components of the two active scalar fields $\Phi_1$ and $\Phi_2$. These VEVs determine the dynamics of the Spontaneous Electro-Weak Symmetry Breaking (EWSB) mechanism of the model, thereby generating masses for the gauge bosons and fermions (as appropriate). The $\eta$ scalar doublet, however, remains inert due to the imposition of the discrete $\mathbb{Z}_2$ symmetry. This symmetry prevents any mixing between the (pseudo)scalar states of $\eta$ and those of $\Phi_1$ and $\Phi_2$, thus ensuring that $\eta$ does not acquire a VEV or contribute to EWSB in the same way as the other doublets.

Consequently, the physical scalar spectrum arising from the two active Higgs doublets in this model is similar to the one in the pure 2HDM. This spectrum consists of four distinct Higgs particles: $h$ (neutral and CP-even), $H$ (neutral and CP-even), $A$ (neutral and CP-odd) and $H^{\pm}$ (charged Higgs bosons), with masses $m_{h}, m_{H}, m_{A}$ and $m_{H^{\pm}}$, respectively \cite{Branco:2011iw,Gunion:2002zf}. In what follows, we shall refer to the lighter CP-even state with a $95$ GeV mass, $h$, as $h_{95}$ and to the heavier CP-even $H$ state with a $125$ GeV mass as $h_{125}$, the former being the candidate to explain the corresponding data anomalies and latter being identified with the SM-like Higgs boson observed in experiments. Furthemore, the $A$ state is the state that we propose being behind the 650 GeV anomaly (hence, it will have a mass around such a value). 

As for the inert sector, the I(1+2)HDM introduces three additional scalar fields: $\chi$, $\chi_{a}$ (which are neutral) and $\chi^{\pm}$ (which is charged). Furthermore, we adopt the {\it dark democracy} approach 
of Refs.~\cite{Keus:2014jha,Keus:2015xya,Cordero-Cid:2016krd,Cordero:2017owj} in what follows, in order to reduce the number of free parameters, thus making the numerical analysis more tractable while preserving the essential features of the I(1+2)HDM. Consequently, by setting $\lambda_a=\lambda_{1133}=\lambda_{2233}$, $\lambda_b=\lambda_{1331}=\lambda_{2332}$ and $\lambda_c=\lambda_{1313}=\lambda_{2323}$, the inert squared masses can be expressed as:
\begin{eqnarray}
&& m^2_{\chi^{\pm}} = m_{\eta}^2 + \frac{1}{2} \lambda_{a} v^2, \nonumber\\
&& m^2_{\chi} = m^2_{\chi^{\pm}} + \frac{1}{2} (\lambda_{b} + \lambda_{c})v^2,\nonumber\\
&& m^2_{\chi_a} = m^2_{\chi^{\pm}} + \frac{1}{2} (\lambda_{b} - \lambda_{c})v^2.
\end{eqnarray}

\noindent
Thus, the Higgs sector of the I(1+2)HDM is described by 12 free parameters:
\begin{align}
\Sigma = \{\, & m_h,\, m_A,\, m_H,\, m_{H^\pm},\, m_{12}^2,\, \tan\beta,\, \sin(\beta-\alpha), \notag \\
             & m_{\chi},\, m_{\chi_a},\, m_{\chi^\pm},\, m_{\eta}^2,\, \lambda_{\eta} \,\}.
\label{set}
\end{align}
Here, $\alpha$ denotes the rotation angle in the CP-even sector, which governs the mixing between the CP-even neutral Higgs states \(h\) and \(H\) whereas the parameter $\beta$ (the mixing angle for both CP-odd and charged sectors) is defined as  $\tan\beta=v_2/v_1$.

After enforcing the symmetries outlined above, the most general Yukawa interactions in the I(1+2)HDM resemble those of the 2HDM  and the corresponding Lagrangian is given by:   
\begin{align*}
\mathcal{L}_{\text{Yukawa}} 
&= - \overline{Q}_L Y_u \widetilde{\Phi}_u u_R 
   - \overline{Q}_L Y_d \Phi_d d_R 
   - \overline{L}_L Y_\ell \Phi_\ell \ell_R + \text{h.c.} \\
&\hspace{-0.9cm}\supset - \sum_{f = u, d, \ell} \left[ 
  \frac{m_f}{v} \kappa^f_h \overline{f} f h 
  + \frac{m_f}{v} \kappa^f_H \overline{f} f H 
  - i \frac{m_f}{v} \kappa^f_A \overline{f} \gamma_5 f A 
\right],
\end{align*}
where the $Y_{f}$'s ($f=u$, $d$ or $l$) are $3\times 3$ Yukawa matrices and $\tilde{\Phi}_{1,2}=i\sigma_{2}\Phi_{1,2}^*$, with $\sigma_{2}$ being the Pauli matrix.
In Tab.~\ref{table1} we list all the Type-I Yukawa reduced couplings $\kappa_h^f $ and $\kappa_H^f $ of the CP-even Higgs bosons, $h$ and $H$.

\begin{table}[ht]
\vspace{0.15cm}
\centering
\large
\begin{tabular}{cccccc}
\hline\hline                        
$\kappa_h^u$  &  $\kappa_h^d$    &  $\kappa_h^\ell$   &  $\kappa_H^u$   &   $\kappa_H^d$  &   $\kappa_H^\ell$  \\ [0.5ex]  
\hline 
$c_\alpha/s_\beta$  &  $c_\alpha/s_\beta$    &   $c_\alpha/s_\beta$   &  $s_\alpha/s_\beta$   &   $s_\alpha/s_\beta$  &   $s_\alpha/s_\beta$ \\
\hline
\end{tabular}
\caption{The I(1+2)HDM Type-I Yukawa couplings  of the neutral Higgs bosons $h,\,H$ 
to the up-quarks, down-quarks and leptons in the I(1+2)HDM, normalised by the corresponding SM Yukawa couplings.}
\label{table1}   
\end{table}

\subsection{Theoretical and Experimental Constraints}
\label{sec:modelconstraints}
In our random parameter scan, we retained only those points that are physically viable, i.e. consistent with both theoretical and experimental requirements. Below we provide a summary of the constraints applied to the I(1+2)HDM Type-I (see \cite{Hmissou:2025uep} for more details):\\[1ex]
$\bullet$ Theoretical Constraints
\begin{itemize}
\item [$\star$] Perturbativity \cite{Moretti:2015cwa}.
\item [$\star$] Unitarity \cite{Moretti:2015cwa}.
\item [$\star$] Vacuum Stability \cite{Grzadkowski:2009bt}.
\end{itemize}
$\bullet$ Experimental Constraints
\begin{itemize}
\item [$\star$] Higgs boson signal strength measurements from the LHC, implemented via \texttt{HiggsSignals-3} \cite{Bechtle:2020uwn}.
\item [$\star$] Exclusion limits on non-SM-like Higgs bosons from direct searches at LEP, the Tevatron, and LHC, as implemented in \texttt{HiggsBounds-6} \cite{Bechtle:2020pkv}.
\item [$\star$] Constraints on the invisible Higgs decay width (in the presence of the inert sector), as automatically included within HiggsTools \cite{Bahl:2022igd}.
\item [$\star$] EW precision tests from $S$, $T$ and $U$ parameters \cite{Merchand:2019bod}, derived following the general result of \cite{Peskin:1991sw,Grimus:2008nb}.
\item [$\star$] DM searches via {\tt micrOMEGAs} \cite{Alguero:2023zol}.
\item [$\star$] Flavour constraints as evaluated using the public {\tt SuperIso} \cite{Mahmoudi:2008tp} program, specifically, several Branching Ratios (BRs) have been checked: $\text{BR}(B \to X_s \gamma)$ \cite{HFLAV:2016hnz}, $\text{BR}(B_s \to \mu^+\mu^-)$ \cite{LHCb:2021awg,LHCb:2021vsc,CMS:2022mgd}, $\text{BR}(B \to \tau\nu)$\cite{HFLAV:2016hnz}, among others. 
\end{itemize}

\subsection{The $\gamma\gamma b \bar{b}$ anomaly}
\label{sec:process}
Here we consider the scenario where the CP-odd scalar $A$, with a mass around 650~GeV, decays via $A \to h_{125} Z$, where the $h_{125}$ subsequently decays as $h_{125} \to \gamma\gamma$, and the $Z$ boson decays as $Z \to b\bar{b}$. For the latter decay, we have incorporated the effect of a lower cut on the invariant mass of the $Z$ decay products ($m_{b\bar b}>70$ GeV) implemented in the CMS analysis of \cite{CMS:2023boe} through the appropriate integration of the Breit-Wigner distribution over the $b\bar{b}$ final state.  This interpretation choice is both theoretically consistent and phenomenologically relevant. On one hand, the original CMS analysis~\cite{CMS:2023boe} modelled the excess in terms of a decay of the form $X_{650} \to h_{125} Y$,  where $Y$ denotes a generic spin-0 resonance. On the other hand, their event selection does not rely on spin or polarisation-sensitise observables. Instead, it targets invariant mass reconstruction and global event kinematics, yet, the phenomenological difference between a pseudoscalar decay and that of a (dominant) longitudinal vector boson is expected to have minimal impact on acceptance. 

With all this in mind,  we also explore other possible channels that would manifest themselves over the I(1+2)HDM parameter space explaining all aforementioned anomalies, 
such as $A \to h_{95} Z$ and $A \to t\bar{t}$, which are indeed competitive with the decay $A \to h_{125} Z$. In particular, we will demonstrate in the following that there exist viable regions in parameter space where the BRs for these two channels are substantial and consistent with all  current  constraints. 

\section{Results and discussions}
\label{sec:res-disc}
Following the previous discussion, we briefly recapitulate the impact of the  applied constraints on the parameter space and present our interpretation of the observed $\gamma\gamma b \bar{b}$ excess within the I(1+2)HDM Type-I. To this end, we perform a numerical scan over the free parameters in Eq.~(\ref{set}), as summarised in Tab.~\ref{tab:par-scan}. The resulting parameter points are subsequently tested against theoretical conditions and experimental data using a dedicated {\tt Python} interface to the aforementioned  packages or own routines to test the most recent limits.
\begin{table}[!t]
\centering
\begin{tabular}{lccccc}
\toprule[1pt]
Parameter & & $m_h$ & $m_H$ & $m_A$ & $m_{H^\pm}$ \\
\midrule[1pt]
Scan Range &  & $[94,\,97]$ & $125.09$  &  $[600,\,700]$   &  $[90,\,10^3]$  \\
\bottomrule[1pt]
\toprule[1pt]
Parameter & & $m_{12}^2$ & $\tan{\beta}$ & $s_{\beta - \alpha}$ &  \\
\midrule[1pt]
Scan Range &  & $[-10^3 ,\, 10^3]$   &   $[0.5,\,25]$  &  $[-0.4 ,\, 0.1]$   &    \\
\bottomrule[1pt]
\toprule[1pt]
Parameter &  & $m_{\eta}^2$ & $m_{\chi},m_{\chi_a},m_{\chi^{\pm}}$ &  $\lambda_{\eta}$ &   \\
\midrule[1pt]
Scan Range &  & $[-10^5 ,\, 10^5]$  &  $[90 ,\, 10^3]$  & $[7 ,\, 12]$  &    \\
\bottomrule[1pt]
\end{tabular}
\caption{The scan ranges of the  I(1+2)HDM Type-I parameters. (Mass (squared) are in GeV$^{(2)}$.)}
\label{tab:par-scan}
\end{table}

Hereafter, for convenience, we shall use the shorthand notation for the relevant cross section:
\begin{equation}
\label{eq:sigma}
\sigma_{\gamma\gamma b \bar{b}} = \sigma(pp \to A \to h_{125} Z \to \gamma\gamma b \bar{b})
\end{equation}

\begin{widetext}
\begin{figure*}
\centering
\includegraphics[width=0.88\columnwidth]{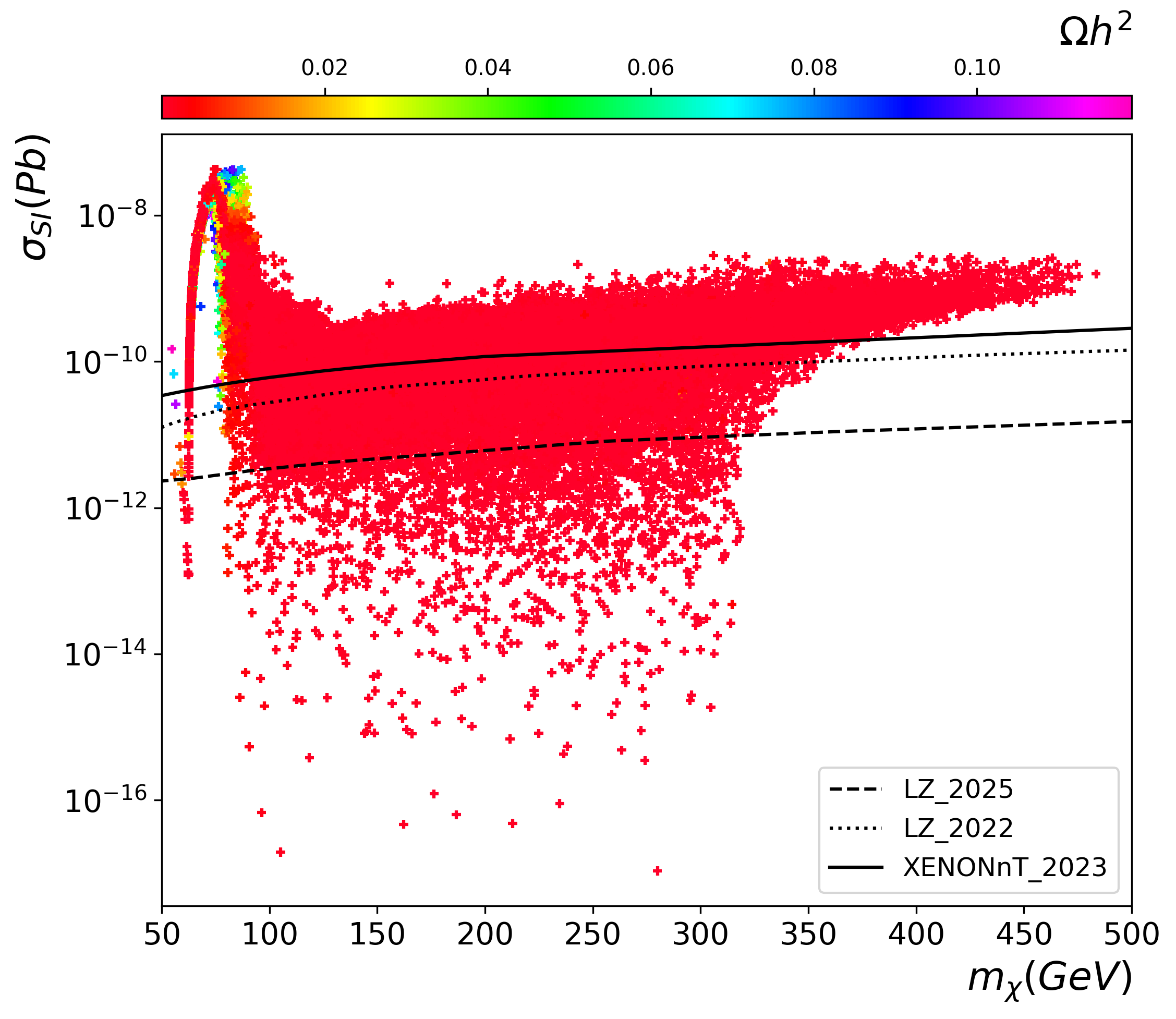}
\includegraphics[width=0.88\columnwidth]{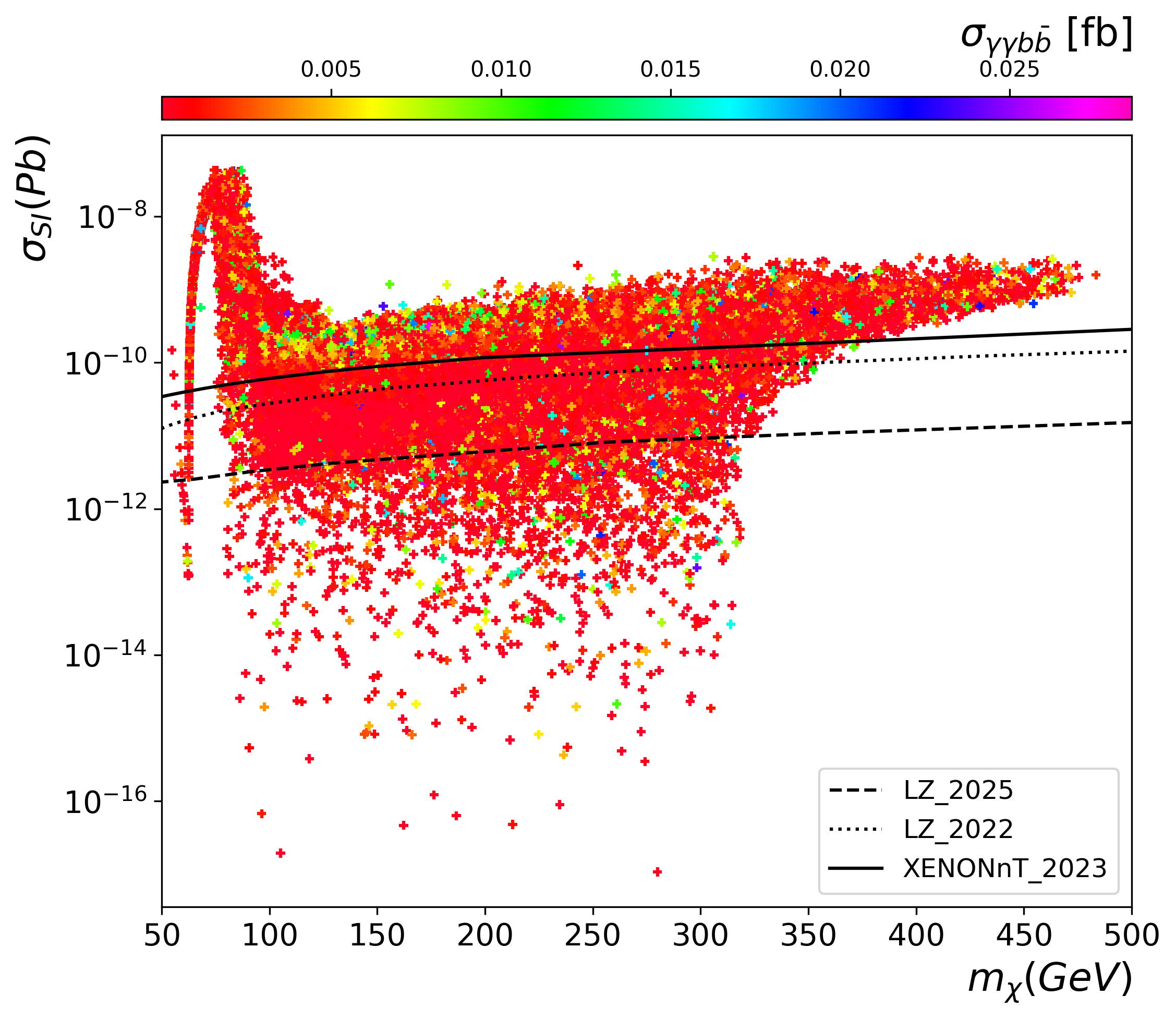}
\caption{The relic density $\Omega h^2$ (left) and the cross section $\sigma_{\gamma\gamma b \bar{b}}$ (right) shown as a scatter plot in the $(m_{\chi},\,\sigma_{SI})$ plane. The XENONnT \cite{XENON:2023cxc} and LUX-ZEPLIN \cite{LZ:2022lsv,LZ:2024zvo} upper limits are shown with
solid, dotted and dashed lines respectively.}
\label{fig0}
\end{figure*}
\end{widetext}

\begin{figure*}
\includegraphics[width=0.329\textwidth]{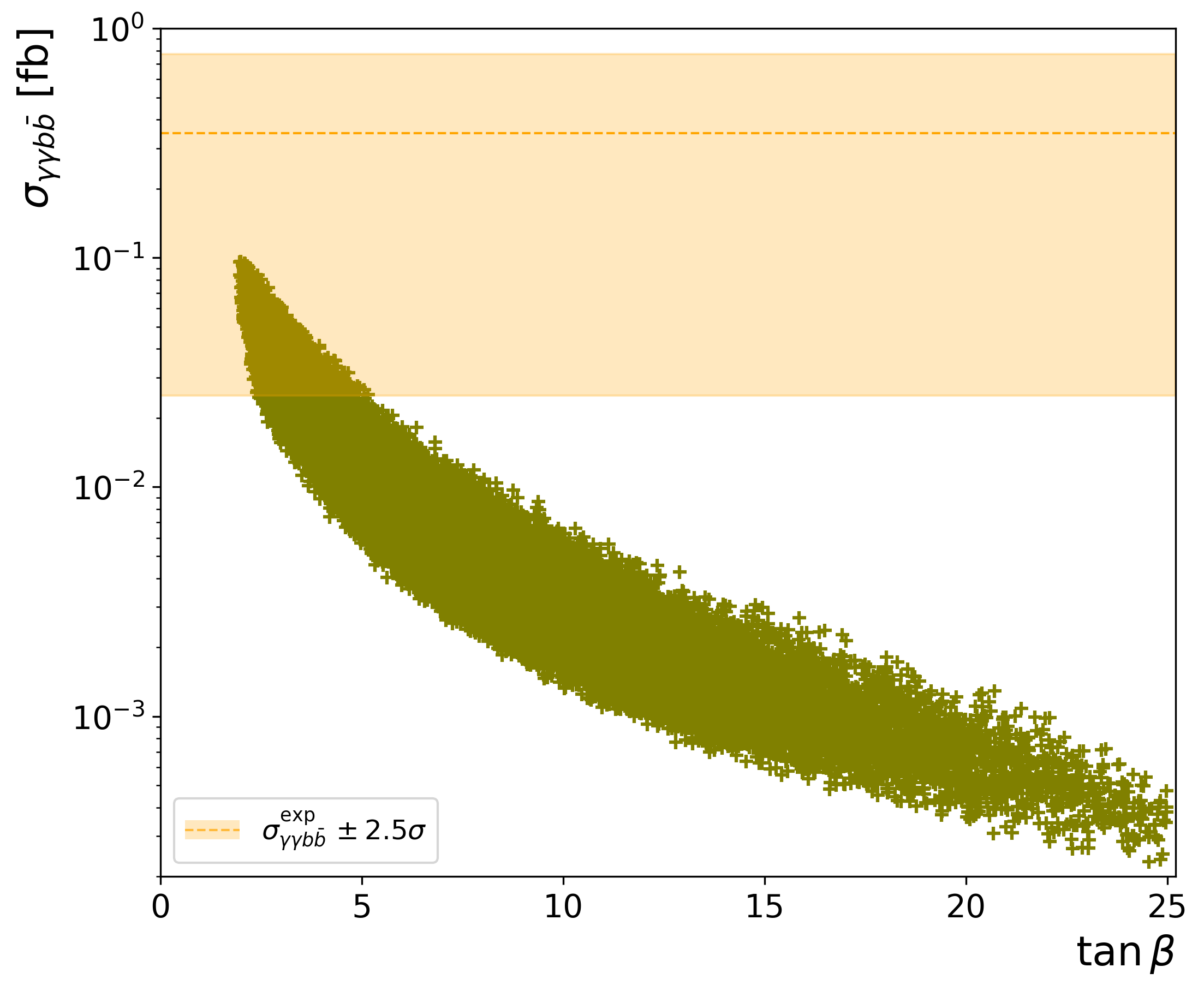}
\includegraphics[width=0.329\textwidth]{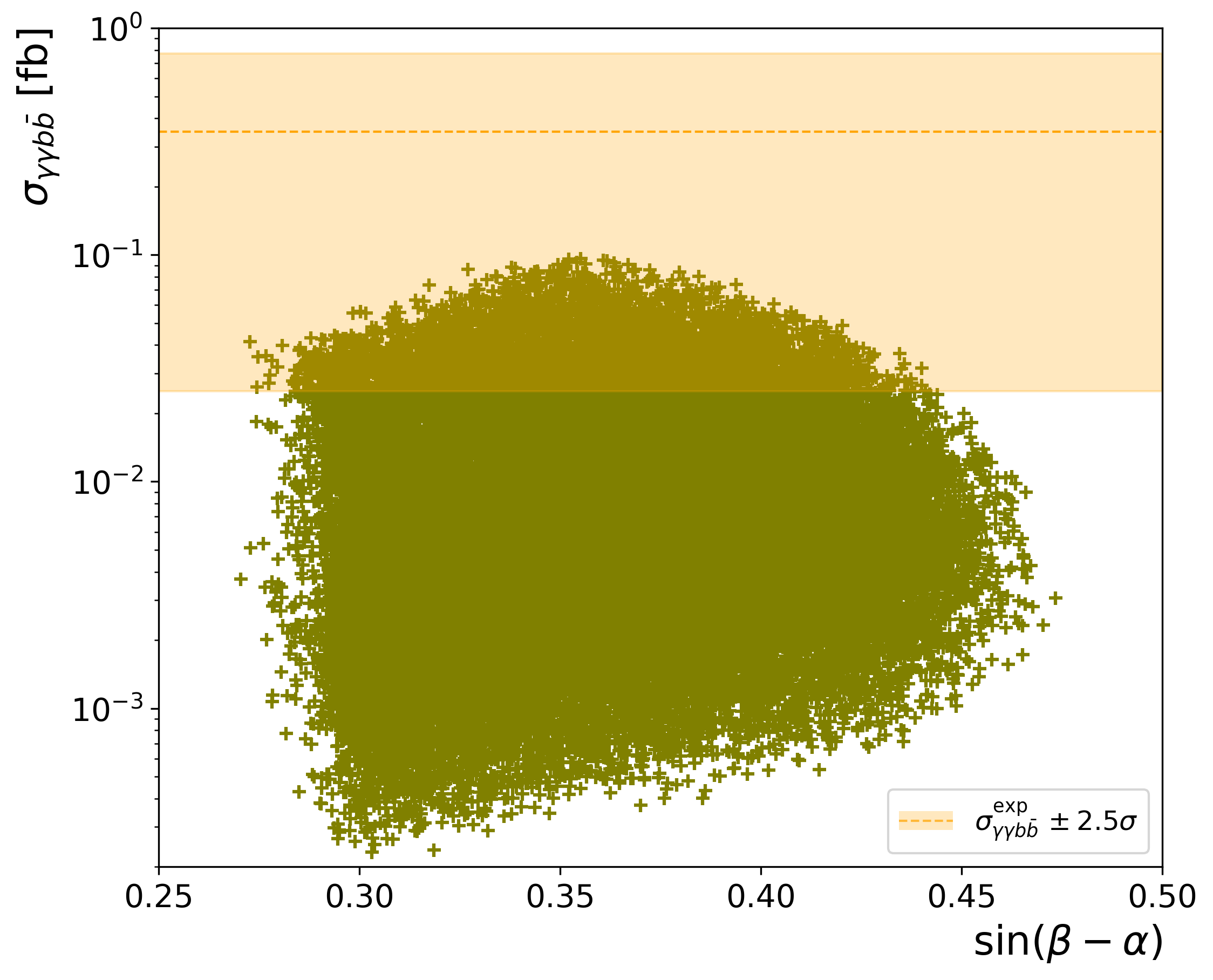}
\includegraphics[width=0.329\textwidth]{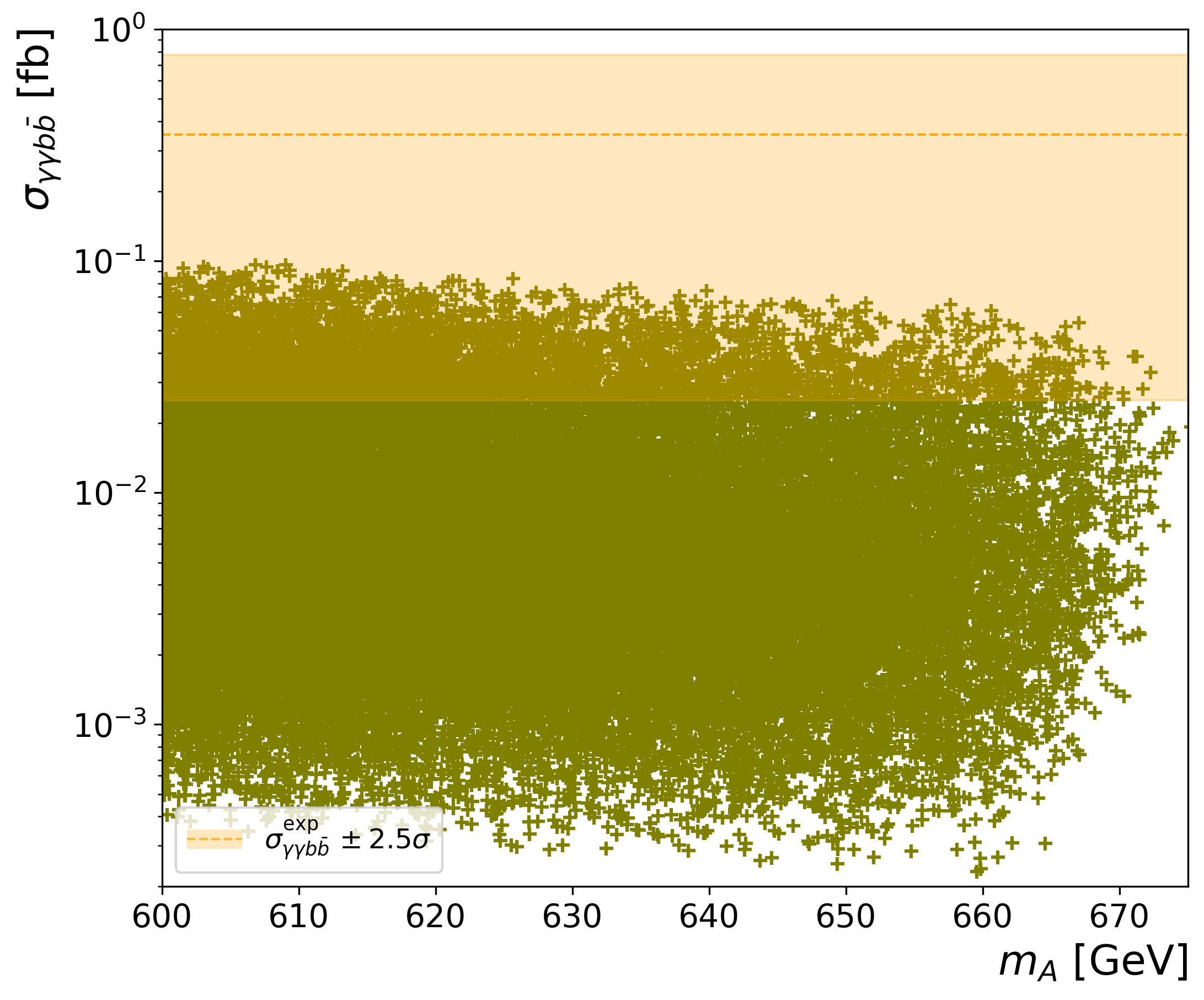}
\caption{The $\sigma_{\gamma\gamma b\bar b}$ values as function of $\tan\beta$ (left),  $\sin({\beta-\alpha})$ (middle) and $m_A$ (right). The orange horizontal band corresponds to the value of the $\gamma\gamma b\bar b$ cross section consistent with the experimental measurement  within $[-2.5\sigma,\,0.5\sigma]$.}
\label{fig1}
\end{figure*}

To start with, as mentioned above, we have explicitly used {\tt microOMEGAs} to  compute the relic density, $\Omega h^2$, and the (in)direct detection rates, the most constraining experiments in the latter respect being XENONnT \cite{XENON:2023cxc} and LUX-ZEPLIN \cite{LZ:2022lsv,LZ:2024zvo}, by taking into account a rescaling of the relevant cross sections in such a way that the computed Spin-Independent ($SI$) DM-nucleon one  $\sigma_{SI}$ reads as
\begin{equation}
\label{eq:spin-indep-xc}
\sigma_{SI}  \to \sigma_{SI} \times \frac{\Omega h^2}{\Omega_{c} h^2},
\end{equation}
where $\Omega_{c} h^2=0.1200 \pm 0.0012$ is the DM relic density determined by the Planck collaboration \cite{Planck:2018vyg}.

Fig.~\ref{fig0} displays $\sigma_{SI}$ as a function of the the DM mass $m_{\chi}$ with the map coding indicating the corresponding DM relic density, $\Omega h^2$ (left), and the values of the $\gamma\gamma b\bar{b}$ signal rate, $\sigma_{\gamma\gamma b\bar{b}}$ (right). As can be seen from the left panel, all the sampled points remain below the Planck limit value while a broad region of parameter space simultaneously satisfies the latest direct detection bounds from XENONnT-2023, LZ-2022 and the projected LZ-2025. Additionally, the right panel in Fig.~\ref{fig0}, except for very low DM masses ($m_{\chi} \le 60\,\text{GeV}$), where most points are excluded, confirms that the DM candidate in the I(1+2)HDM not only survives the most stringent constraints from direct detection experiments but also yields values of $\sigma_{\gamma\gamma b\bar{b}}$ consistent with the CMS excess at 650 GeV at $-2.5\sigma$ for larger masses ($100\,\text{GeV} \le m_{\chi} \le 350\,\text{GeV}$). Thus, the DM and collider sectors of the model are closely correlated. Most importantly, here,  the relic density is not oversaturated and both direct and indirect constraints are satisfied.

Subsequently, we found it appropriate to show the possible dependencies of such an observable on the model parameters. As can be seen from Fig.~\ref{fig1} (left),  a low $\tan\beta$  $\le 5$ is necessary to accommodate the $\sigma_{\gamma\gamma b\bar b}$ value corresponding to the excess. Additionally, the range of $\sin(\beta-\alpha)$ that matches the $\gamma\gamma b\bar b$ cross section value up to $-2.5\sigma$ is relatively wide, i.e.,  $0.27-0.47$ (see Fig.~\ref{fig1} (middle)). 

\begin{figure*}[!ht]
\centering
\includegraphics[width=0.329\textwidth]{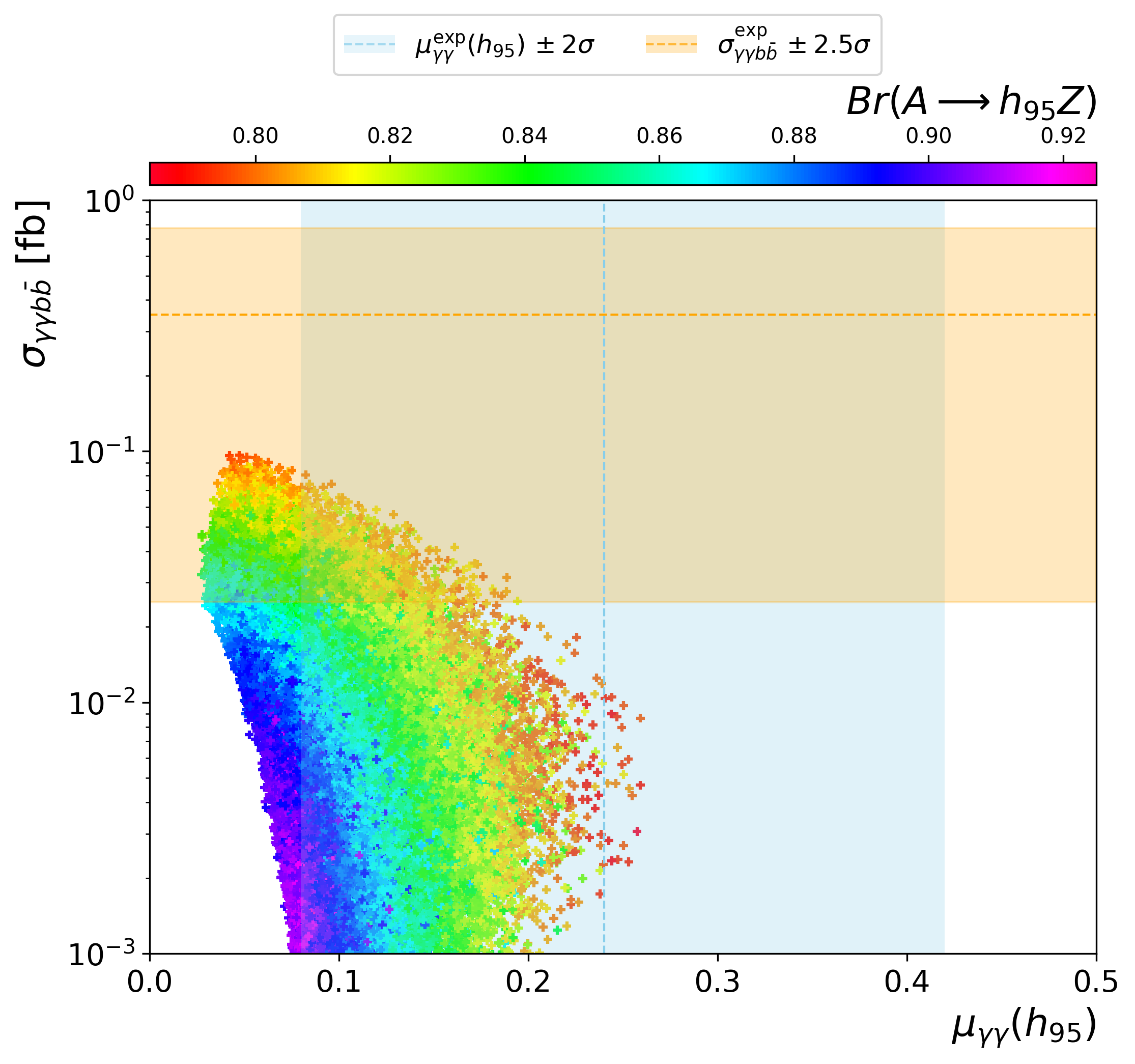}
\includegraphics[width=0.329\textwidth]{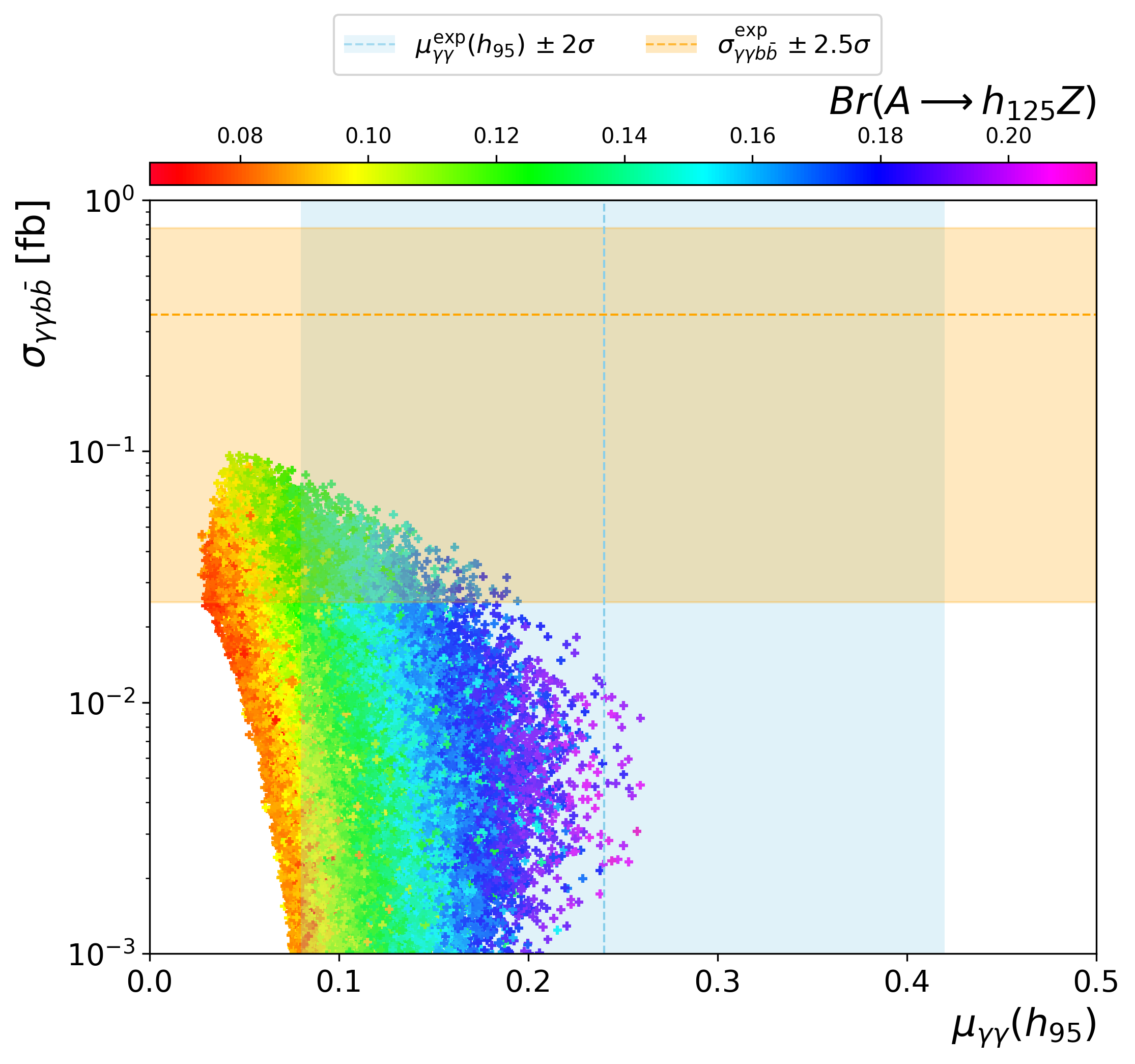}
\includegraphics[width=0.329\textwidth]{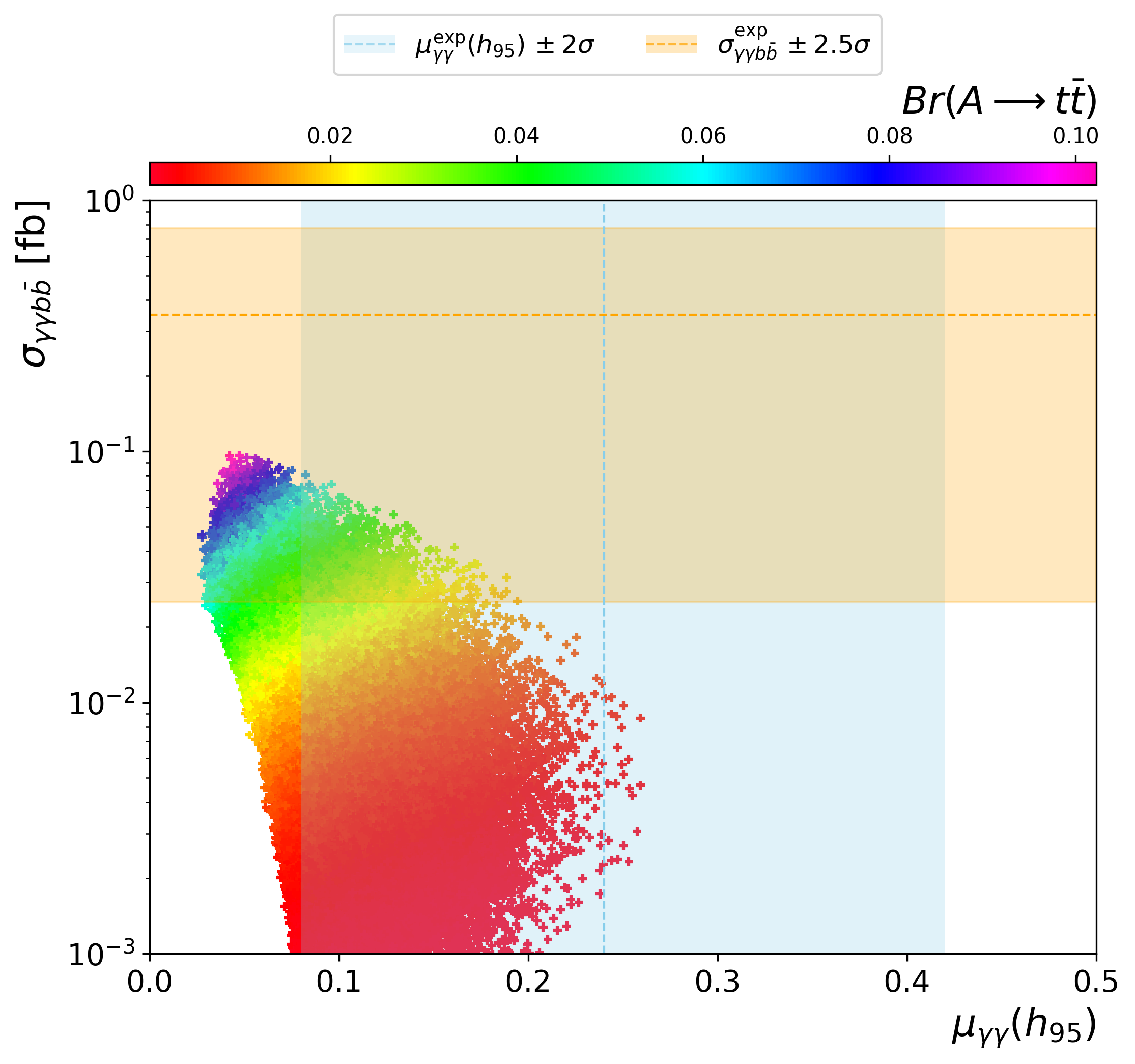}\\
\includegraphics[width=0.329\textwidth]{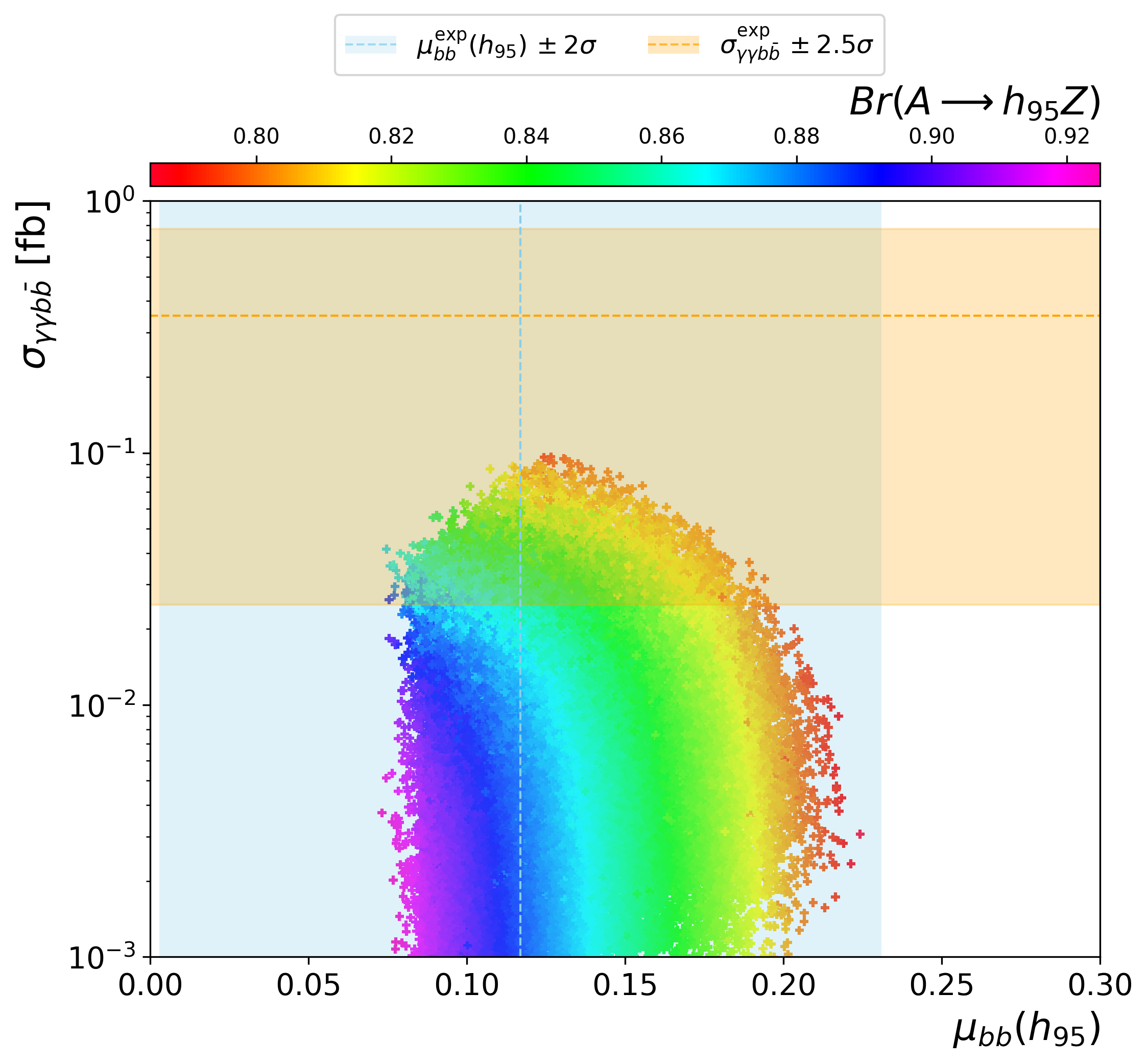}
\includegraphics[width=0.329\textwidth]{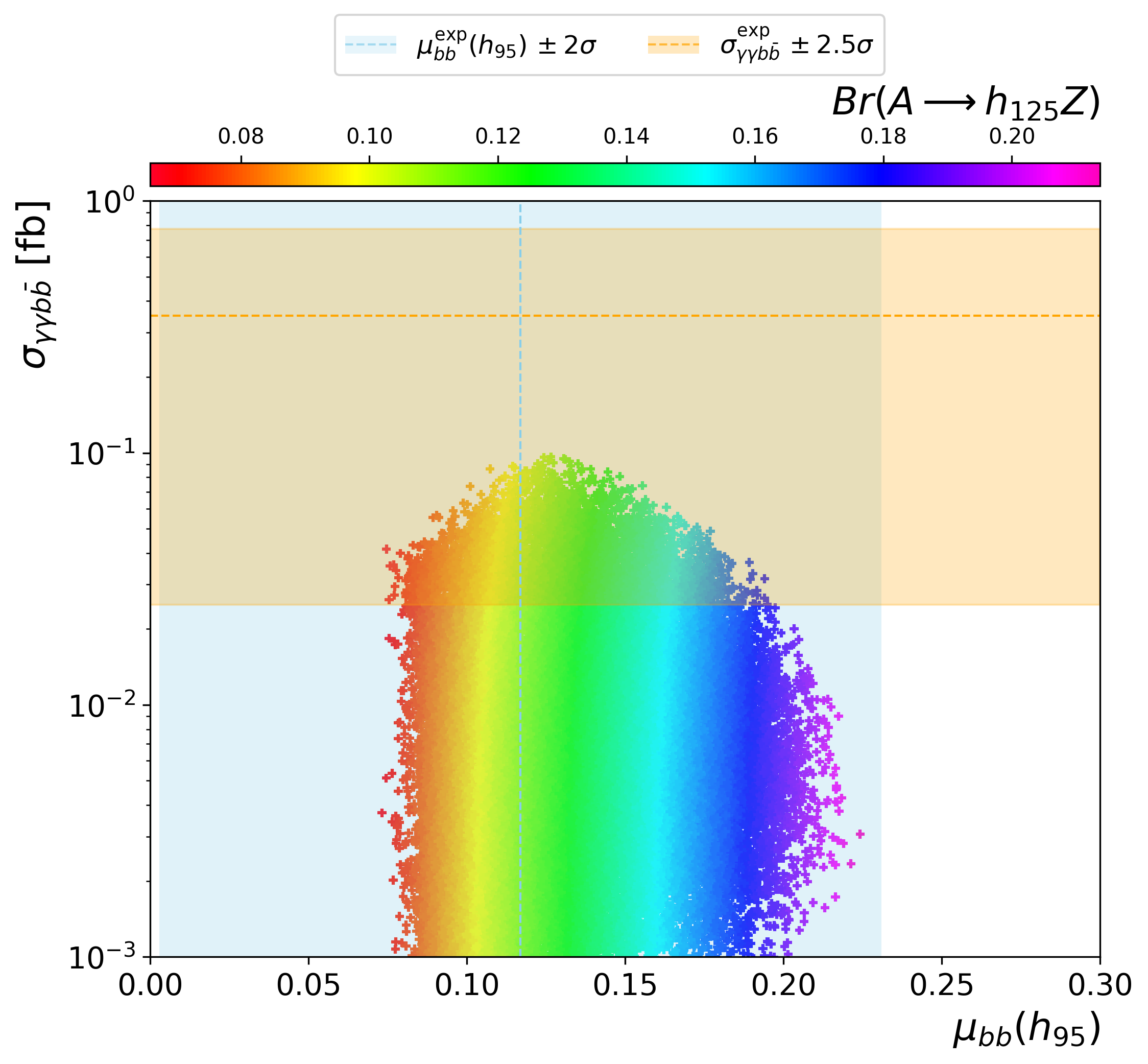}
\includegraphics[width=0.329\textwidth]{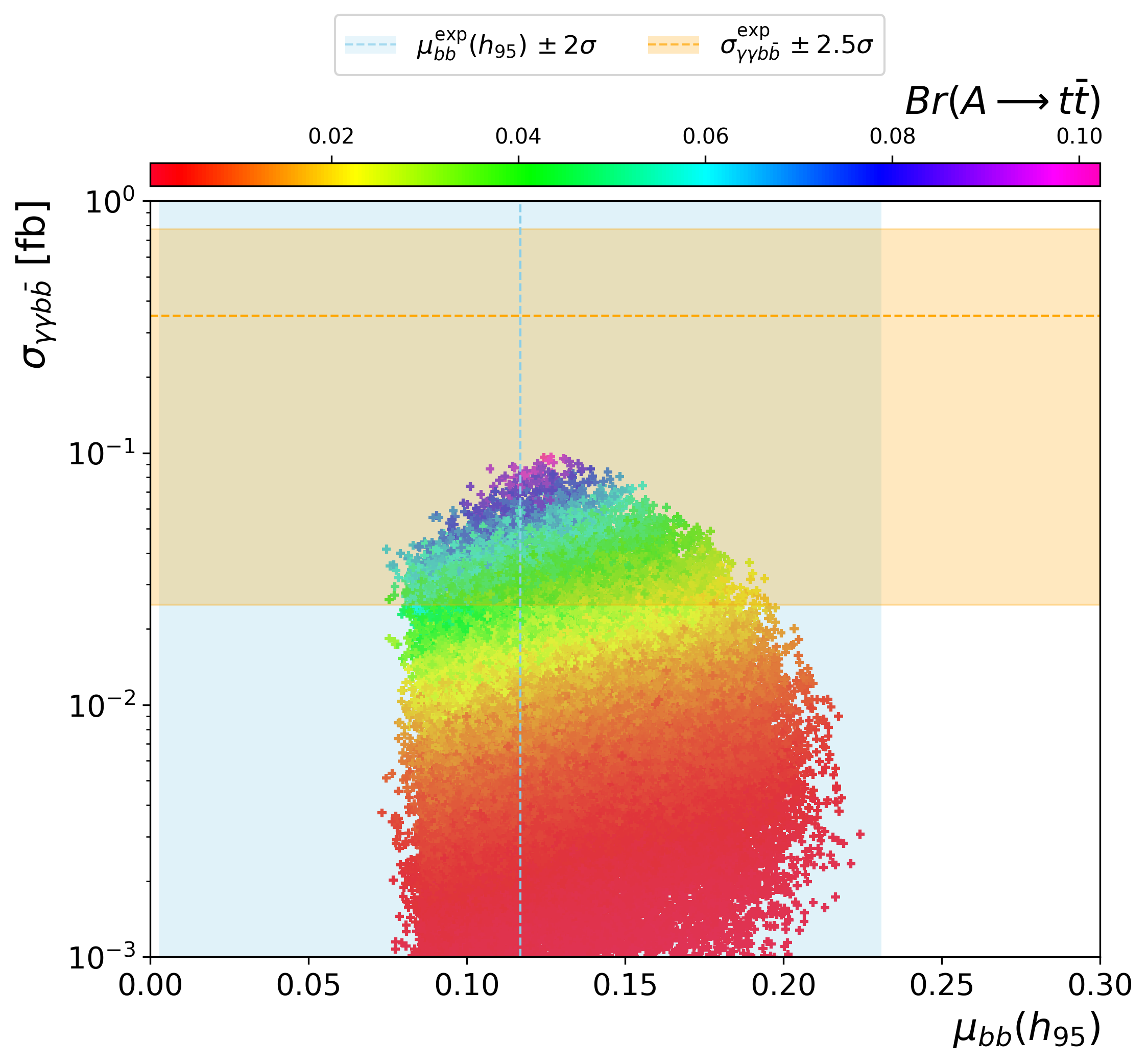}
\caption{The scatter plots of the production cross section $\sigma_{\gamma\gamma b \bar{b}}$, as a function of the signal strengths $\mu_{\gamma\gamma}(h_{95})$ (top) and $\mu_{b\bar{b}}(h_{95})$ (bottom). The colour coding indicates one of the following BRs: $\text{BR}(A \to h_{95})$ (left), $\text{BR}(A \to h_{125} Z)$ (middle) and $\text{BR}(A \to t\bar{t})$ (rigtht). Furthermore,  in orange(cyan) horizontal(vertical) band represents the measured value of $\sigma_{\gamma\gamma b \bar{b}}$ (signal strengths $\mu_{\gamma\gamma}(h_{95})$ and $\mu_{b\bar{b}}(h)$).}
\label{fig2}
\end{figure*}

Regarding the right panel in Fig.~\ref{fig1}, it shows that the $\sigma_{\gamma\gamma b \bar{b}}$ decreases with $m_{A}$ happens for phase space reasons, so that larger values of $m_{A}$ require larger values of the BR$(h_{125} \to \gamma\gamma)$. Unambiguously, though, our analysis indicates that the full mass range of $m_A$ between 600 and 700 GeV remains viable in order to account for the observed excess.

In this connection, it should be noted here that inert model parameters, while not explicitly entering the discussion, do affect the phenomenology of this process indirectly, in particular, the additional scalar charged states with mass $100 \le m_{\chi^\pm} \le 500$ GeV alter the di-photon decay width of the SM-like Higgs boson,  ultimately leading to $\sigma_{\gamma\gamma b \bar{b}}$ value close to the measured one. At the same time, they enter the $h_{95}\to\gamma\gamma$ decay process, enabling the I(1+2)HDM studied here to explain the 95 GeV excesses, as explained in  \cite{Hmissou:2025uep}. 

We then show in Fig.~\ref{fig2} the correlations between the predicted values for $\sigma_{\gamma\gamma b \bar{b}}$ and the signal strengths $\mu_{\gamma\gamma}(h_{95})$ and $\mu_{b\bar b}(h_{95})$, as defined in  \cite{Hmissou:2025uep}, 
of the lighter CP-even Higgs boson $h_{95}$, for different values of BR$(A \to h_{95}\,Z)$, BR$(A \to h_{125}\,Z)$ and BR$(A \to t\bar{t})$ (ignoring the signal strength $\mu_{\tau^+\tau^-}(h_{95})$, as the excess is here more marginal in comparison).

The figure highlights the $[-2.5\sigma,\,0.5\sigma]$ horizontal region in orange corresponding to the $\gamma\gamma b\bar b$ anomaly (as in Fig.~\ref{fig1}) and, as a vertical band depicted in cyan, the range of values of the $\mu_{\gamma\gamma}$ (upper panel) and $\mu_{b\bar{b}}$ (lower panel) excesses, both drawn at the $\pm2\sigma$ level. From this figure, it can be seen that it is in general quite difficult to simultaneously accommodate the  $\gamma\gamma b \bar{b}$ excess with the $\mu_{\gamma\gamma}(h_{95})$ and $\mu_{b\bar{b}}(h_{95})$ ones, but a non-zero overlapping region nevertheless exists. 

Quite interestingly, the points most compatible with the CMS excess at 650 GeV lie within or near the experimentally allowed regions, particularly in the low-to moderate di-photon signal strength, i.e., $\mu_{\gamma\gamma}(h_{95}) \lesssim 0.2$, see Fig.~\ref{fig2} 
(top frames), and remain consistent with the $\mu_{b\bar b}(h_{95})$ one at $2\sigma$ level, see Fig.~\ref{fig2} (lower frames). For the values of BRs shown, the maximum value for $\sigma_{\gamma\gamma b \bar{b}}\approx0.096$ fb is reached for BR$(A \to h_{95}\,Z)\approx79.16\%$, BR$(A \to h_{125}\,Z)\approx10.57\%$ and BR$(A \to t\bar{t})\approx9.86\%$. Overall, we see that, for fixed $\mu_{\gamma\gamma}(h_{95})$ and/or $\mu_{b\bar b}(h_{95})$, $\sigma_{\gamma\gamma b \bar{b}}$ increases inversely with BR$(A \to h_{95}\,Z)$, revealing that this decay mode dominates in the viable parameter space, with the largest BRs observed where the signal strengths $\mu_{b\bar{b}}(h_{95})$ and $\mu_{\gamma\gamma}(h_{95})$ remain consistent with experimental  bounds, so that this is the key probe of the BSM scenario adopted here.  
Furthermore, notice that  BR$(A \to t\bar{t})$ is generally not negligible, so that this channel may also be of relevance for testing the viability of the I(1+2)HDM Type-I against all the studied datasets. All this is  essentially exemplifying  the fact that the $A \rightarrow h_{125} Z$ mode (i.e., the apparently anomalous one) 
 is not the only available channel for probing our scenario, so the present 650 GeV anomaly may actually be accompanied by others.

\begin{figure}[!ht]
\centering
\includegraphics[width=0.9\columnwidth]{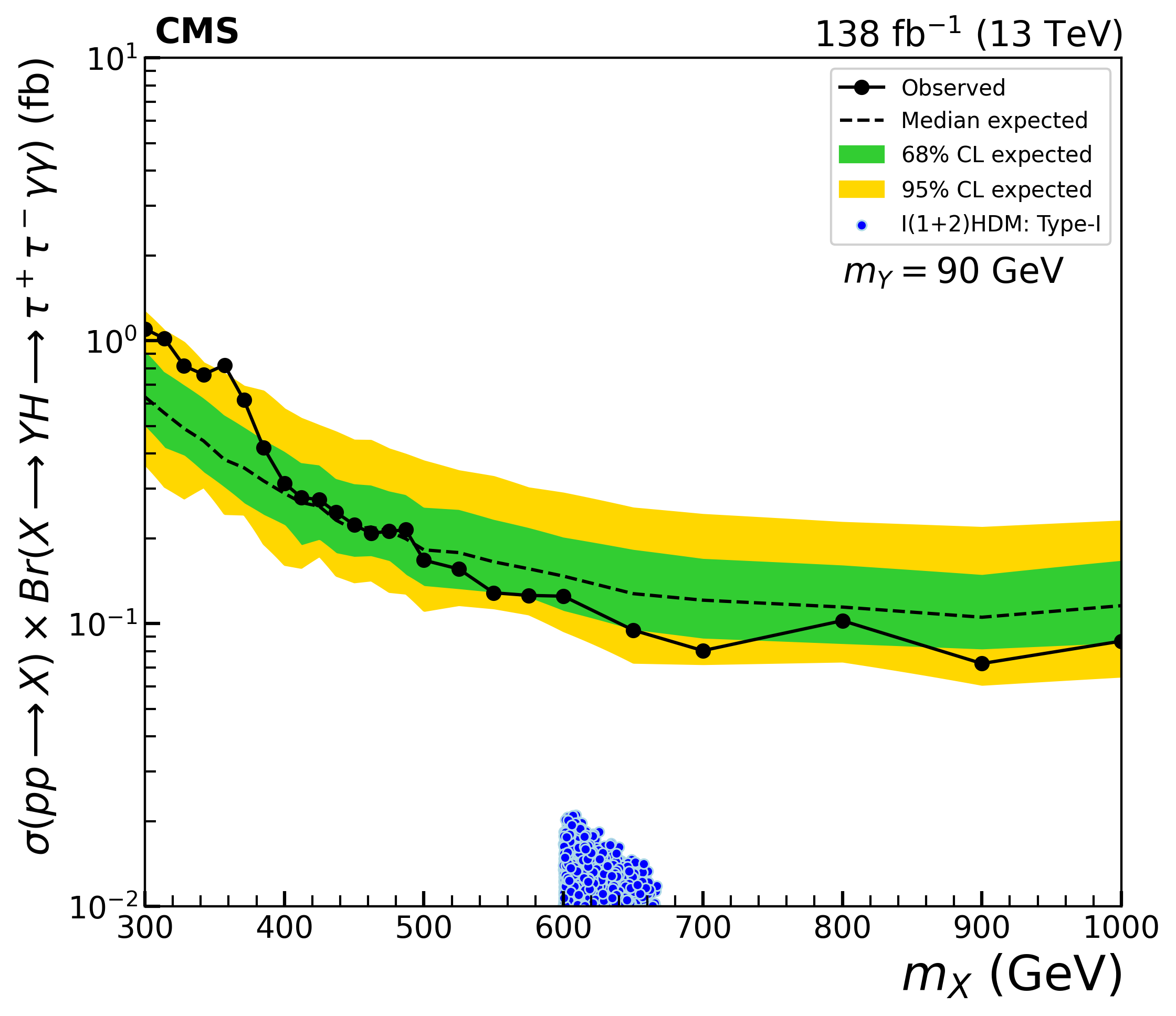}
\caption{The $\sigma({pp\to X\to HY\to \gamma\gamma \tau^+\tau^-})$ values as function of $m_X$ as obtained for the I(1+2)HDM points explaining the 650 and 95 GeV anomalies. Notice that, in such a scenario, one has $X=A$, $Y=Z$ and $H=h_{125}$.}
\label{fig3}
\end{figure}

As a final point, and as already intimated, 
 we note that the CMS Collaboration has recently carried out a dedicated analysis focusing on a possible heavy spin-0 resonance $X$ decaying into the SM-like Higgs boson  and a lighter object $Y$ (again, potentially but not necessarily a spin-0 one), with the distinctive final state $\gamma\gamma\tau^+\tau^-$ \cite{CMS:2025tqi}. They studied in particular the mass regions $m_X\approx 650$ GeV and $m_Y\approx 95$ GeV and found no significant excess.
This search is thus relevant for our case as  it  places stringent upper limits on the production cross section times BR for the channel $pp\to A \to h_{125}Z \to \gamma\gamma \tau^+\tau^-$ in our model, ranging from 0.69 to 15 fb as a function of the $X$ mass. We have therefore accounted for this in our analysis. As shown in Fig.~\ref{fig3}, our predictions for this process, over the region of I(1+2)HDM Type-I parameter space explaining the 650 and 95 GeV anomalies simultaneously, 
 lie below the current sensitivity lines, yet not exceedingly so. Hence, while the $\gamma\gamma\tau^+\tau^-$ final state remains unconstrained presently in our model it also represents a promising probe of the I(1+2)HDM Type-I for upcoming searches at  Run 3 and the HL-LHC \cite{Gianotti:2002xx}.

\section{Conclusions}
In this paper, we have tested the I(1+2)HDM Type-I as a possible theoretical scenario explaining a variety of anomalies that have recently appeared in experimental data, which origin could well be in an extended Higgs sector, with respect to the one embedded in the SM. Specifically, excesses have been seen in the search for new Higgs bosons with masses both below and above the one of the SM-like Higgs state detected at the LHC in 2012, with a mass of 125 GeV. In fact, anomalies have appeared around 95 GeV in analyses targeting the following processes: $e^+e^-\to Z^{\ast} \to  h_{95} Z \to  b\bar b + {\rm jets}$ at LEP and $pp\to h_{95}\to \gamma\gamma$ and $\tau^+\tau^-$ at the LHC. Furthermore, another quite significant excess has been extracted at the LHC around 650 GeV, potentially emerging from the process $pp\to A\to  h_{125}Z \to \gamma\gamma  b\bar b$ (with $m_A={650}$ GeV). All such production and decay channels are present in the I(1+2)HDM and, over sizable regions of the parameter space of such a BSM scenario, they can explain all aforementioned excesses at approximately the 2$\sigma$ CL (possibly apart from the di-tau excess, which is quite marginal, in fact). Crucially, this is made possible thanks to the presence of additional scalar charged states, a feature specific to our BSM setup (not existing, e.g., in more conventional (N)2HDMs), as they enable a perfect mapping of the 95 GeV anomaly in the di-photon channel, the most constraining one by far amongst all the discussed datasets. 

In addition, alongside explaining existing anomalies, hence, through a data analysis done {\sl a posteriori}, we have also proposed, over the parameter space of the I(1+2)HDM accommodating the various excesses, two collateral signals which should eventually appear in data. These should then be considered as {\sl a priori} manifestations that one should expect from our theoretical framework for it to be the one underpinning Nature. The new signatures are the processes $pp\to A\to h_{95}Z$ and $pp\to A\to t\bar t$, which can be searched for in a variety of final states. 
We thus  encourage the experimental Collaborations working at the current LHC and/or future HL-LHC to test such predictions, which constitute the hallmark 
manifestations of the I(1+2)HDM. For this purpose, we will make available upon request all our data.

\section*{Acknowledgments}
SM is supported in part through the NExT Institute and  STFC Consolidated Grant No. ST/L000296/1. LR would like to thank the CERN Department of Theoretical Physics for its hospitality and stimulating environment, where a part of this work was carried out.

\bibliography{ref}

\begin{thebibliography}{70}%
\makeatletter
\providecommand \@ifxundefined [1]{%
 \@ifx{#1\undefined}
}%
\providecommand \@ifnum [1]{%
 \ifnum #1\expandafter \@firstoftwo
 \else \expandafter \@secondoftwo
 \fi
}%
\providecommand \@ifx [1]{%
 \ifx #1\expandafter \@firstoftwo
 \else \expandafter \@secondoftwo
 \fi
}%
\providecommand \natexlab [1]{#1}%
\providecommand \enquote  [1]{``#1''}%
\providecommand \bibnamefont  [1]{#1}%
\providecommand \bibfnamefont [1]{#1}%
\providecommand \citenamefont [1]{#1}%
\providecommand \href@noop [0]{\@secondoftwo}%
\providecommand \href [0]{\begingroup \@sanitize@url \@href}%
\providecommand \@href[1]{\@@startlink{#1}\@@href}%
\providecommand \@@href[1]{\endgroup#1\@@endlink}%
\providecommand \@sanitize@url [0]{\catcode `\\12\catcode `\$12\catcode
  `\&12\catcode `\#12\catcode `\^12\catcode `\_12\catcode `\%12\relax}%
\providecommand \@@startlink[1]{}%
\providecommand \@@endlink[0]{}%
\providecommand \url  [0]{\begingroup\@sanitize@url \@url }%
\providecommand \@url [1]{\endgroup\@href {#1}{\urlprefix }}%
\providecommand \urlprefix  [0]{URL }%
\providecommand \Eprint [0]{\href }%
\providecommand \doibase [0]{https://doi.org/}%
\providecommand \selectlanguage [0]{\@gobble}%
\providecommand \bibinfo  [0]{\@secondoftwo}%
\providecommand \bibfield  [0]{\@secondoftwo}%
\providecommand \translation [1]{[#1]}%
\providecommand \BibitemOpen [0]{}%
\providecommand \bibitemStop [0]{}%
\providecommand \bibitemNoStop [0]{.\EOS\space}%
\providecommand \EOS [0]{\spacefactor3000\relax}%
\providecommand \BibitemShut  [1]{\csname bibitem#1\endcsname}%
\let\auto@bib@innerbib\@empty
\bibitem [{\citenamefont {Aad}\ \emph {et~al.}(2012)\citenamefont {Aad} \emph
  {et~al.}}]{ATLAS:2012yve}%
  \BibitemOpen
  \bibfield  {author} {\bibinfo {author} {\bibfnamefont {G.}~\bibnamefont
  {Aad}} \emph {et~al.} (\bibinfo {collaboration} {ATLAS}),\ }\bibfield
  {title} {\bibinfo {title} {{Observation of a new particle in the search for
  the Standard Model Higgs boson with the ATLAS detector at the LHC}},\ }\href
  {https://doi.org/10.1016/j.physletb.2012.08.020} {\bibfield  {journal}
  {\bibinfo  {journal} {Phys. Lett. B}\ }\textbf {\bibinfo {volume} {716}},\
  \bibinfo {pages} {1} (\bibinfo {year} {2012})},\ \Eprint
  {https://arxiv.org/abs/1207.7214} {arXiv:1207.7214 [hep-ex]} \BibitemShut
  {NoStop}%
\bibitem [{\citenamefont {Chatrchyan}\ \emph {et~al.}(2012)\citenamefont
  {Chatrchyan} \emph {et~al.}}]{CMS:2012qbp}%
  \BibitemOpen
  \bibfield  {author} {\bibinfo {author} {\bibfnamefont {S.}~\bibnamefont
  {Chatrchyan}} \emph {et~al.} (\bibinfo {collaboration} {CMS}),\ }\bibfield
  {title} {\bibinfo {title} {{Observation of a New Boson at a Mass of 125 GeV
  with the CMS Experiment at the LHC}},\ }\href
  {https://doi.org/10.1016/j.physletb.2012.08.021} {\bibfield  {journal}
  {\bibinfo  {journal} {Phys. Lett. B}\ }\textbf {\bibinfo {volume} {716}},\
  \bibinfo {pages} {30} (\bibinfo {year} {2012})},\ \Eprint
  {https://arxiv.org/abs/1207.7235} {arXiv:1207.7235 [hep-ex]} \BibitemShut
  {NoStop}%
\bibitem [{\citenamefont {Barate}\ \emph {et~al.}(2003)\citenamefont {Barate}
  \emph {et~al.}}]{LEPWorkingGroupforHiggsbosonsearches:2003ing}%
  \BibitemOpen
  \bibfield  {author} {\bibinfo {author} {\bibfnamefont {R.}~\bibnamefont
  {Barate}} \emph {et~al.} (\bibinfo {collaboration} {LEP Working Group for
  Higgs boson searches, ALEPH, DELPHI, L3, OPAL}),\ }\bibfield  {title}
  {\bibinfo {title} {{Search for the standard model Higgs boson at LEP}},\
  }\href {https://doi.org/10.1016/S0370-2693(03)00614-2} {\bibfield  {journal}
  {\bibinfo  {journal} {Phys. Lett. B}\ }\textbf {\bibinfo {volume} {565}},\
  \bibinfo {pages} {61} (\bibinfo {year} {2003})},\ \Eprint
  {https://arxiv.org/abs/hep-ex/0306033} {arXiv:hep-ex/0306033} \BibitemShut
  {NoStop}%
\bibitem [{\citenamefont {Schael}\ \emph {et~al.}(2006)\citenamefont {Schael}
  \emph {et~al.}}]{ALEPH:2006tnd}%
  \BibitemOpen
  \bibfield  {author} {\bibinfo {author} {\bibfnamefont {S.}~\bibnamefont
  {Schael}} \emph {et~al.} (\bibinfo {collaboration} {ALEPH, DELPHI, L3, OPAL,
  LEP Working Group for Higgs Boson Searches}),\ }\bibfield  {title} {\bibinfo
  {title} {{Search for neutral MSSM Higgs bosons at LEP}},\ }\href
  {https://doi.org/10.1140/epjc/s2006-02569-7} {\bibfield  {journal} {\bibinfo
  {journal} {Eur. Phys. J. C}\ }\textbf {\bibinfo {volume} {47}},\ \bibinfo
  {pages} {547} (\bibinfo {year} {2006})},\ \Eprint
  {https://arxiv.org/abs/hep-ex/0602042} {arXiv:hep-ex/0602042} \BibitemShut
  {NoStop}%
\bibitem [{\citenamefont {Cao}\ \emph {et~al.}(2017)\citenamefont {Cao},
  \citenamefont {Guo}, \citenamefont {He}, \citenamefont {Wu},\ and\
  \citenamefont {Zhang}}]{Cao:2016uwt}%
  \BibitemOpen
  \bibfield  {author} {\bibinfo {author} {\bibfnamefont {J.}~\bibnamefont
  {Cao}}, \bibinfo {author} {\bibfnamefont {X.}~\bibnamefont {Guo}}, \bibinfo
  {author} {\bibfnamefont {Y.}~\bibnamefont {He}}, \bibinfo {author}
  {\bibfnamefont {P.}~\bibnamefont {Wu}},\ and\ \bibinfo {author}
  {\bibfnamefont {Y.}~\bibnamefont {Zhang}},\ }\bibfield  {title} {\bibinfo
  {title} {{Diphoton signal of the light Higgs boson in natural NMSSM}},\
  }\href {https://doi.org/10.1103/PhysRevD.95.116001} {\bibfield  {journal}
  {\bibinfo  {journal} {Phys. Rev. D}\ }\textbf {\bibinfo {volume} {95}},\
  \bibinfo {pages} {116001} (\bibinfo {year} {2017})},\ \Eprint
  {https://arxiv.org/abs/1612.08522} {arXiv:1612.08522 [hep-ph]} \BibitemShut
  {NoStop}%
\bibitem [{\citenamefont {Hayrapetyan}\ \emph
  {et~al.}(2025{\natexlab{a}})\citenamefont {Hayrapetyan} \emph
  {et~al.}}]{CMS:2024yhz}%
  \BibitemOpen
  \bibfield  {author} {\bibinfo {author} {\bibfnamefont {A.}~\bibnamefont
  {Hayrapetyan}} \emph {et~al.} (\bibinfo {collaboration} {CMS}),\ }\bibfield
  {title} {\bibinfo {title} {{Search for a standard model-like Higgs boson in
  the mass range between 70 and 110 GeV in the diphoton final state in
  proton-proton collisions at s=13TeV}},\ }\href
  {https://doi.org/10.1016/j.physletb.2024.139067} {\bibfield  {journal}
  {\bibinfo  {journal} {Phys. Lett. B}\ }\textbf {\bibinfo {volume} {860}},\
  \bibinfo {pages} {139067} (\bibinfo {year} {2025}{\natexlab{a}})},\ \Eprint
  {https://arxiv.org/abs/2405.18149} {arXiv:2405.18149 [hep-ex]} \BibitemShut
  {NoStop}%
\bibitem [{\citenamefont {Tumasyan}\ \emph
  {et~al.}(2023{\natexlab{a}})\citenamefont {Tumasyan} \emph
  {et~al.}}]{CMS:2022goy}%
  \BibitemOpen
  \bibfield  {author} {\bibinfo {author} {\bibfnamefont {A.}~\bibnamefont
  {Tumasyan}} \emph {et~al.} (\bibinfo {collaboration} {CMS}),\ }\bibfield
  {title} {\bibinfo {title} {{Searches for additional Higgs bosons and for
  vector leptoquarks in $\tau\tau$ final states in proton-proton collisions at
  $\sqrt{s}$ = 13 TeV}},\ }\href {https://doi.org/10.1007/JHEP07(2023)073}
  {\bibfield  {journal} {\bibinfo  {journal} {JHEP}\ }\textbf {\bibinfo
  {volume} {07}},\ \bibinfo {pages} {073}},\ \Eprint
  {https://arxiv.org/abs/2208.02717} {arXiv:2208.02717 [hep-ex]} \BibitemShut
  {NoStop}%
\bibitem [{ATL(2018)}]{ATLAS:2018xad}%
  \BibitemOpen
  \bibfield  {title} {\bibinfo {title} {{Search for resonances in the 65 to 110
  GeV diphoton invariant mass range using 80 fb$^{-1}$ of $pp$ collisions
  collected at $\sqrt{s}=13$ TeV with the ATLAS detector}},\ }\href@noop {} {\
  (\bibinfo {year} {2018})}\BibitemShut {NoStop}%
\bibitem [{Note1()}]{Note1}%
  \BibitemOpen
  \bibinfo {note} {Note that the best-fit mass actually differs from one
  channel to another, however, given the poor resolution in invariant mass of
  $b\protect \bar b$ and $\tau ^+\tau ^-$ pairs, compared with to the $\gamma
  \gamma $ case (which is centered at 95 or so GeV), we collectively use the
  latter value throughout, including in our forthcoming tests of the various
  anomalies.}\BibitemShut {Stop}%
\bibitem [{\citenamefont {Tumasyan}\ \emph {et~al.}(2024)\citenamefont
  {Tumasyan} \emph {et~al.}}]{CMS:2023boe}%
  \BibitemOpen
  \bibfield  {author} {\bibinfo {author} {\bibfnamefont {A.}~\bibnamefont
  {Tumasyan}} \emph {et~al.} (\bibinfo {collaboration} {CMS}),\ }\bibfield
  {title} {\bibinfo {title} {{Search for a new resonance decaying into two
  spin-0 bosons in a final state with two photons and two bottom quarks in
  proton-proton collisions at $ \sqrt{s} $ = 13 TeV}},\ }\href
  {https://doi.org/10.1007/JHEP05(2024)316} {\bibfield  {journal} {\bibinfo
  {journal} {JHEP}\ }\textbf {\bibinfo {volume} {05}},\ \bibinfo {pages}
  {316}},\ \Eprint {https://arxiv.org/abs/2310.01643} {arXiv:2310.01643
  [hep-ex]} \BibitemShut {NoStop}%
\bibitem [{\citenamefont {Cacciapaglia}\ \emph {et~al.}(2016)\citenamefont
  {Cacciapaglia}, \citenamefont {Deandrea}, \citenamefont {Gascon-Shotkin},
  \citenamefont {Le~Corre}, \citenamefont {Lethuillier},\ and\ \citenamefont
  {Tao}}]{Cacciapaglia:2016tlr}%
  \BibitemOpen
  \bibfield  {author} {\bibinfo {author} {\bibfnamefont {G.}~\bibnamefont
  {Cacciapaglia}}, \bibinfo {author} {\bibfnamefont {A.}~\bibnamefont
  {Deandrea}}, \bibinfo {author} {\bibfnamefont {S.}~\bibnamefont
  {Gascon-Shotkin}}, \bibinfo {author} {\bibfnamefont {S.}~\bibnamefont
  {Le~Corre}}, \bibinfo {author} {\bibfnamefont {M.}~\bibnamefont
  {Lethuillier}},\ and\ \bibinfo {author} {\bibfnamefont {J.}~\bibnamefont
  {Tao}},\ }\bibfield  {title} {\bibinfo {title} {{Search for a lighter Higgs
  boson in Two Higgs Doublet Models}},\ }\href
  {https://doi.org/10.1007/JHEP12(2016)068} {\bibfield  {journal} {\bibinfo
  {journal} {JHEP}\ }\textbf {\bibinfo {volume} {12}},\ \bibinfo {pages}
  {068}},\ \Eprint {https://arxiv.org/abs/1607.08653} {arXiv:1607.08653
  [hep-ph]} \BibitemShut {NoStop}%
\bibitem [{\citenamefont {Benbrik}\ \emph
  {et~al.}(2022{\natexlab{a}})\citenamefont {Benbrik}, \citenamefont {Boukidi},
  \citenamefont {Moretti},\ and\ \citenamefont {Semlali}}]{Benbrik:2022azi}%
  \BibitemOpen
  \bibfield  {author} {\bibinfo {author} {\bibfnamefont {R.}~\bibnamefont
  {Benbrik}}, \bibinfo {author} {\bibfnamefont {M.}~\bibnamefont {Boukidi}},
  \bibinfo {author} {\bibfnamefont {S.}~\bibnamefont {Moretti}},\ and\ \bibinfo
  {author} {\bibfnamefont {S.}~\bibnamefont {Semlali}},\ }\bibfield  {title}
  {\bibinfo {title} {{Explaining the 96 GeV Di-photon anomaly in a generic 2HDM
  Type-III}},\ }\href {https://doi.org/10.1016/j.physletb.2022.137245}
  {\bibfield  {journal} {\bibinfo  {journal} {Phys. Lett. B}\ }\textbf
  {\bibinfo {volume} {832}},\ \bibinfo {pages} {137245} (\bibinfo {year}
  {2022}{\natexlab{a}})},\ \Eprint {https://arxiv.org/abs/2204.07470}
  {arXiv:2204.07470 [hep-ph]} \BibitemShut {NoStop}%
\bibitem [{\citenamefont {Benbrik}\ \emph
  {et~al.}(2022{\natexlab{b}})\citenamefont {Benbrik}, \citenamefont {Boukidi},
  \citenamefont {Moretti},\ and\ \citenamefont {Semlali}}]{Benbrik:2022tlg}%
  \BibitemOpen
  \bibfield  {author} {\bibinfo {author} {\bibfnamefont {R.}~\bibnamefont
  {Benbrik}}, \bibinfo {author} {\bibfnamefont {M.}~\bibnamefont {Boukidi}},
  \bibinfo {author} {\bibfnamefont {S.}~\bibnamefont {Moretti}},\ and\ \bibinfo
  {author} {\bibfnamefont {S.}~\bibnamefont {Semlali}},\ }\bibfield  {title}
  {\bibinfo {title} {{Probing a 96 GeV Higgs Boson in the Di-Photon Channel at
  the LHC}},\ }\href {https://doi.org/10.22323/1.414.0547} {\bibfield
  {journal} {\bibinfo  {journal} {PoS}\ }\textbf {\bibinfo {volume}
  {ICHEP2022}},\ \bibinfo {pages} {547} (\bibinfo {year}
  {2022}{\natexlab{b}})},\ \Eprint {https://arxiv.org/abs/2211.11140}
  {arXiv:2211.11140 [hep-ph]} \BibitemShut {NoStop}%
\bibitem [{\citenamefont {Khanna}\ \emph {et~al.}(2024)\citenamefont {Khanna},
  \citenamefont {Moretti},\ and\ \citenamefont {Sarkar}}]{Khanna:2024bah}%
  \BibitemOpen
  \bibfield  {author} {\bibinfo {author} {\bibfnamefont {A.}~\bibnamefont
  {Khanna}}, \bibinfo {author} {\bibfnamefont {S.}~\bibnamefont {Moretti}},\
  and\ \bibinfo {author} {\bibfnamefont {A.}~\bibnamefont {Sarkar}},\
  }\bibfield  {title} {\bibinfo {title} {{Explaining 95 (or so) GeV Anomalies
  in the 2-Higgs Doublet Model Type-I}},\ }\href@noop {} {\  (\bibinfo {year}
  {2024})},\ \Eprint {https://arxiv.org/abs/2409.02587} {arXiv:2409.02587
  [hep-ph]} \BibitemShut {NoStop}%
\bibitem [{\citenamefont {Biek\"otter}\ \emph {et~al.}(2020)\citenamefont
  {Biek\"otter}, \citenamefont {Chakraborti},\ and\ \citenamefont
  {Heinemeyer}}]{Biekotter:2019kde}%
  \BibitemOpen
  \bibfield  {author} {\bibinfo {author} {\bibfnamefont {T.}~\bibnamefont
  {Biek\"otter}}, \bibinfo {author} {\bibfnamefont {M.}~\bibnamefont
  {Chakraborti}},\ and\ \bibinfo {author} {\bibfnamefont {S.}~\bibnamefont
  {Heinemeyer}},\ }\bibfield  {title} {\bibinfo {title} {{A 96 GeV Higgs boson
  in the N2HDM}},\ }\href {https://doi.org/10.1140/epjc/s10052-019-7561-2}
  {\bibfield  {journal} {\bibinfo  {journal} {Eur. Phys. J. C}\ }\textbf
  {\bibinfo {volume} {80}},\ \bibinfo {pages} {2} (\bibinfo {year} {2020})},\
  \Eprint {https://arxiv.org/abs/1903.11661} {arXiv:1903.11661 [hep-ph]}
  \BibitemShut {NoStop}%
\bibitem [{\citenamefont {Heinemeyer}\ \emph {et~al.}(2022)\citenamefont
  {Heinemeyer}, \citenamefont {Li}, \citenamefont {Lika}, \citenamefont
  {Moortgat-Pick},\ and\ \citenamefont {Paasch}}]{Heinemeyer:2021msz}%
  \BibitemOpen
  \bibfield  {author} {\bibinfo {author} {\bibfnamefont {S.}~\bibnamefont
  {Heinemeyer}}, \bibinfo {author} {\bibfnamefont {C.}~\bibnamefont {Li}},
  \bibinfo {author} {\bibfnamefont {F.}~\bibnamefont {Lika}}, \bibinfo {author}
  {\bibfnamefont {G.}~\bibnamefont {Moortgat-Pick}},\ and\ \bibinfo {author}
  {\bibfnamefont {S.}~\bibnamefont {Paasch}},\ }\bibfield  {title} {\bibinfo
  {title} {{Phenomenology of a 96~GeV Higgs boson in the 2HDM with an
  additional singlet}},\ }\href {https://doi.org/10.1103/PhysRevD.106.075003}
  {\bibfield  {journal} {\bibinfo  {journal} {Phys. Rev. D}\ }\textbf {\bibinfo
  {volume} {106}},\ \bibinfo {pages} {075003} (\bibinfo {year} {2022})},\
  \Eprint {https://arxiv.org/abs/2112.11958} {arXiv:2112.11958 [hep-ph]}
  \BibitemShut {NoStop}%
\bibitem [{\citenamefont {Biek\"otter}\ \emph
  {et~al.}(2022{\natexlab{a}})\citenamefont {Biek\"otter}, \citenamefont
  {Heinemeyer},\ and\ \citenamefont {Weiglein}}]{Biekotter:2022jyr}%
  \BibitemOpen
  \bibfield  {author} {\bibinfo {author} {\bibfnamefont {T.}~\bibnamefont
  {Biek\"otter}}, \bibinfo {author} {\bibfnamefont {S.}~\bibnamefont
  {Heinemeyer}},\ and\ \bibinfo {author} {\bibfnamefont {G.}~\bibnamefont
  {Weiglein}},\ }\bibfield  {title} {\bibinfo {title} {{Mounting evidence for a
  95 GeV Higgs boson}},\ }\href {https://doi.org/10.1007/JHEP08(2022)201}
  {\bibfield  {journal} {\bibinfo  {journal} {JHEP}\ }\textbf {\bibinfo
  {volume} {08}},\ \bibinfo {pages} {201}},\ \Eprint
  {https://arxiv.org/abs/2203.13180} {arXiv:2203.13180 [hep-ph]} \BibitemShut
  {NoStop}%
\bibitem [{\citenamefont {Banik}\ \emph {et~al.}(2023)\citenamefont {Banik},
  \citenamefont {Crivellin}, \citenamefont {Iguro},\ and\ \citenamefont
  {Kitahara}}]{Banik:2023ecr}%
  \BibitemOpen
  \bibfield  {author} {\bibinfo {author} {\bibfnamefont {S.}~\bibnamefont
  {Banik}}, \bibinfo {author} {\bibfnamefont {A.}~\bibnamefont {Crivellin}},
  \bibinfo {author} {\bibfnamefont {S.}~\bibnamefont {Iguro}},\ and\ \bibinfo
  {author} {\bibfnamefont {T.}~\bibnamefont {Kitahara}},\ }\bibfield  {title}
  {\bibinfo {title} {{Asymmetric di-Higgs signals of the next-to-minimal 2HDM
  with a U(1) symmetry}},\ }\href {https://doi.org/10.1103/PhysRevD.108.075011}
  {\bibfield  {journal} {\bibinfo  {journal} {Phys. Rev. D}\ }\textbf {\bibinfo
  {volume} {108}},\ \bibinfo {pages} {075011} (\bibinfo {year} {2023})},\
  \Eprint {https://arxiv.org/abs/2303.11351} {arXiv:2303.11351 [hep-ph]}
  \BibitemShut {NoStop}%
\bibitem [{\citenamefont {Biek\"otter}\ \emph {et~al.}(2023)\citenamefont
  {Biek\"otter}, \citenamefont {Heinemeyer},\ and\ \citenamefont
  {Weiglein}}]{Biekotter:2023jld}%
  \BibitemOpen
  \bibfield  {author} {\bibinfo {author} {\bibfnamefont {T.}~\bibnamefont
  {Biek\"otter}}, \bibinfo {author} {\bibfnamefont {S.}~\bibnamefont
  {Heinemeyer}},\ and\ \bibinfo {author} {\bibfnamefont {G.}~\bibnamefont
  {Weiglein}},\ }\bibfield  {title} {\bibinfo {title} {{The CMS di-photon
  excess at 95 GeV in view of the LHC Run 2 results}},\ }\href@noop {} {\
  (\bibinfo {year} {2023})},\ \Eprint {https://arxiv.org/abs/2303.12018}
  {arXiv:2303.12018 [hep-ph]} \BibitemShut {NoStop}%
\bibitem [{\citenamefont {Biek\"otter}\ \emph {et~al.}(2024)\citenamefont
  {Biek\"otter}, \citenamefont {Heinemeyer},\ and\ \citenamefont
  {Weiglein}}]{Biekotter:2023oen}%
  \BibitemOpen
  \bibfield  {author} {\bibinfo {author} {\bibfnamefont {T.}~\bibnamefont
  {Biek\"otter}}, \bibinfo {author} {\bibfnamefont {S.}~\bibnamefont
  {Heinemeyer}},\ and\ \bibinfo {author} {\bibfnamefont {G.}~\bibnamefont
  {Weiglein}},\ }\bibfield  {title} {\bibinfo {title} {{95.4~GeV diphoton
  excess at ATLAS and CMS}},\ }\href
  {https://doi.org/10.1103/PhysRevD.109.035005} {\bibfield  {journal} {\bibinfo
   {journal} {Phys. Rev. D}\ }\textbf {\bibinfo {volume} {109}},\ \bibinfo
  {pages} {035005} (\bibinfo {year} {2024})},\ \Eprint
  {https://arxiv.org/abs/2306.03889} {arXiv:2306.03889 [hep-ph]} \BibitemShut
  {NoStop}%
\bibitem [{\citenamefont {Kundu}\ \emph {et~al.}(2024)\citenamefont {Kundu},
  \citenamefont {Mondal},\ and\ \citenamefont {Moultaka}}]{Kundu:2024sip}%
  \BibitemOpen
  \bibfield  {author} {\bibinfo {author} {\bibfnamefont {A.}~\bibnamefont
  {Kundu}}, \bibinfo {author} {\bibfnamefont {P.}~\bibnamefont {Mondal}},\ and\
  \bibinfo {author} {\bibfnamefont {G.}~\bibnamefont {Moultaka}},\ }\bibfield
  {title} {\bibinfo {title} {{Indications for new scalar resonances at the LHC
  and a possible interpretation}},\ }\href@noop {} {\  (\bibinfo {year}
  {2024})},\ \Eprint {https://arxiv.org/abs/2411.14126} {arXiv:2411.14126
  [hep-ph]} \BibitemShut {NoStop}%
\bibitem [{\citenamefont {Ahriche}(2024)}]{Ahriche:2023wkj}%
  \BibitemOpen
  \bibfield  {author} {\bibinfo {author} {\bibfnamefont {A.}~\bibnamefont
  {Ahriche}},\ }\bibfield  {title} {\bibinfo {title} {{95~GeV excess in the
  Georgi-Machacek model: Single or twin peak resonance}},\ }\href
  {https://doi.org/10.1103/PhysRevD.110.035010} {\bibfield  {journal} {\bibinfo
   {journal} {Phys. Rev. D}\ }\textbf {\bibinfo {volume} {110}},\ \bibinfo
  {pages} {035010} (\bibinfo {year} {2024})},\ \Eprint
  {https://arxiv.org/abs/2312.10484} {arXiv:2312.10484 [hep-ph]} \BibitemShut
  {NoStop}%
\bibitem [{\citenamefont {Chen}\ \emph {et~al.}(2024)\citenamefont {Chen},
  \citenamefont {Chiang}, \citenamefont {Heinemeyer},\ and\ \citenamefont
  {Weiglein}}]{Chen:2023bqr}%
  \BibitemOpen
  \bibfield  {author} {\bibinfo {author} {\bibfnamefont {T.-K.}\ \bibnamefont
  {Chen}}, \bibinfo {author} {\bibfnamefont {C.-W.}\ \bibnamefont {Chiang}},
  \bibinfo {author} {\bibfnamefont {S.}~\bibnamefont {Heinemeyer}},\ and\
  \bibinfo {author} {\bibfnamefont {G.}~\bibnamefont {Weiglein}},\ }\bibfield
  {title} {\bibinfo {title} {{95~GeV Higgs boson in the Georgi-Machacek
  model}},\ }\href {https://doi.org/10.1103/PhysRevD.109.075043} {\bibfield
  {journal} {\bibinfo  {journal} {Phys. Rev. D}\ }\textbf {\bibinfo {volume}
  {109}},\ \bibinfo {pages} {075043} (\bibinfo {year} {2024})},\ \Eprint
  {https://arxiv.org/abs/2312.13239} {arXiv:2312.13239 [hep-ph]} \BibitemShut
  {NoStop}%
\bibitem [{\citenamefont {Du}\ \emph {et~al.}(2025)\citenamefont {Du},
  \citenamefont {Liu},\ and\ \citenamefont {Chang}}]{Du:2025eop}%
  \BibitemOpen
  \bibfield  {author} {\bibinfo {author} {\bibfnamefont {X.}~\bibnamefont
  {Du}}, \bibinfo {author} {\bibfnamefont {H.}~\bibnamefont {Liu}},\ and\
  \bibinfo {author} {\bibfnamefont {Q.}~\bibnamefont {Chang}},\ }\bibfield
  {title} {\bibinfo {title} {{Interpretation of 95 GeV Excess within the
  Georgi-Machacek Model in Light of Positive Definiteness Constraints}},\
  }\href@noop {} {\  (\bibinfo {year} {2025})},\ \Eprint
  {https://arxiv.org/abs/2502.06444} {arXiv:2502.06444 [hep-ph]} \BibitemShut
  {NoStop}%
\bibitem [{\citenamefont {Arhrib}\ \emph {et~al.}(2025)\citenamefont {Arhrib},
  \citenamefont {Phan}, \citenamefont {Tran},\ and\ \citenamefont
  {Yuan}}]{Arhrib:2024zsw}%
  \BibitemOpen
  \bibfield  {author} {\bibinfo {author} {\bibfnamefont {A.}~\bibnamefont
  {Arhrib}}, \bibinfo {author} {\bibfnamefont {K.~H.}\ \bibnamefont {Phan}},
  \bibinfo {author} {\bibfnamefont {V.~Q.}\ \bibnamefont {Tran}},\ and\
  \bibinfo {author} {\bibfnamefont {T.-C.}\ \bibnamefont {Yuan}},\ }\bibfield
  {title} {\bibinfo {title} {{When the Standard Model Higgs meets its lighter
  95 GeV twin}},\ }\href {https://doi.org/10.1016/j.nuclphysb.2025.116909}
  {\bibfield  {journal} {\bibinfo  {journal} {Nucl. Phys. B}\ }\textbf
  {\bibinfo {volume} {1015}},\ \bibinfo {pages} {116909} (\bibinfo {year}
  {2025})},\ \Eprint {https://arxiv.org/abs/2405.03127} {arXiv:2405.03127
  [hep-ph]} \BibitemShut {NoStop}%
\bibitem [{\citenamefont {Hmissou}\ \emph {et~al.}(2025)\citenamefont
  {Hmissou}, \citenamefont {Moretti},\ and\ \citenamefont
  {Rahili}}]{Hmissou:2025uep}%
  \BibitemOpen
  \bibfield  {author} {\bibinfo {author} {\bibfnamefont {A.}~\bibnamefont
  {Hmissou}}, \bibinfo {author} {\bibfnamefont {S.}~\bibnamefont {Moretti}},\
  and\ \bibinfo {author} {\bibfnamefont {L.}~\bibnamefont {Rahili}},\
  }\bibfield  {title} {\bibinfo {title} {{Investigating the 95 GeV Higgs Boson
  Excesses within the I(1+2)HDM}},\ }\href@noop {} {\  (\bibinfo {year}
  {2025})},\ \Eprint {https://arxiv.org/abs/2502.03631} {arXiv:2502.03631
  [hep-ph]} \BibitemShut {NoStop}%
\bibitem [{\citenamefont {Cao}\ \emph {et~al.}(2020)\citenamefont {Cao},
  \citenamefont {Jia}, \citenamefont {Yue}, \citenamefont {Zhou},\ and\
  \citenamefont {Zhu}}]{Cao:2019ofo}%
  \BibitemOpen
  \bibfield  {author} {\bibinfo {author} {\bibfnamefont {J.}~\bibnamefont
  {Cao}}, \bibinfo {author} {\bibfnamefont {X.}~\bibnamefont {Jia}}, \bibinfo
  {author} {\bibfnamefont {Y.}~\bibnamefont {Yue}}, \bibinfo {author}
  {\bibfnamefont {H.}~\bibnamefont {Zhou}},\ and\ \bibinfo {author}
  {\bibfnamefont {P.}~\bibnamefont {Zhu}},\ }\bibfield  {title} {\bibinfo
  {title} {{96 GeV diphoton excess in seesaw extensions of the natural
  NMSSM}},\ }\href {https://doi.org/10.1103/PhysRevD.101.055008} {\bibfield
  {journal} {\bibinfo  {journal} {Phys. Rev. D}\ }\textbf {\bibinfo {volume}
  {101}},\ \bibinfo {pages} {055008} (\bibinfo {year} {2020})},\ \Eprint
  {https://arxiv.org/abs/1908.07206} {arXiv:1908.07206 [hep-ph]} \BibitemShut
  {NoStop}%
\bibitem [{\citenamefont {Biek\"otter}\ \emph
  {et~al.}(2022{\natexlab{b}})\citenamefont {Biek\"otter}, \citenamefont
  {Grohsjean}, \citenamefont {Heinemeyer}, \citenamefont {Schwanenberger},\
  and\ \citenamefont {Weiglein}}]{Biekotter:2021qbc}%
  \BibitemOpen
  \bibfield  {author} {\bibinfo {author} {\bibfnamefont {T.}~\bibnamefont
  {Biek\"otter}}, \bibinfo {author} {\bibfnamefont {A.}~\bibnamefont
  {Grohsjean}}, \bibinfo {author} {\bibfnamefont {S.}~\bibnamefont
  {Heinemeyer}}, \bibinfo {author} {\bibfnamefont {C.}~\bibnamefont
  {Schwanenberger}},\ and\ \bibinfo {author} {\bibfnamefont {G.}~\bibnamefont
  {Weiglein}},\ }\bibfield  {title} {\bibinfo {title} {{Possible indications
  for new Higgs bosons in the reach of the LHC: N2HDM and NMSSM
  interpretations}},\ }\href {https://doi.org/10.1140/epjc/s10052-022-10099-1}
  {\bibfield  {journal} {\bibinfo  {journal} {Eur. Phys. J. C}\ }\textbf
  {\bibinfo {volume} {82}},\ \bibinfo {pages} {178} (\bibinfo {year}
  {2022}{\natexlab{b}})},\ \Eprint {https://arxiv.org/abs/2109.01128}
  {arXiv:2109.01128 [hep-ph]} \BibitemShut {NoStop}%
\bibitem [{\citenamefont {Li}\ \emph {et~al.}(2023)\citenamefont {Li},
  \citenamefont {Qiao}, \citenamefont {Wang},\ and\ \citenamefont
  {Zhu}}]{Li:2023kbf}%
  \BibitemOpen
  \bibfield  {author} {\bibinfo {author} {\bibfnamefont {W.}~\bibnamefont
  {Li}}, \bibinfo {author} {\bibfnamefont {H.}~\bibnamefont {Qiao}}, \bibinfo
  {author} {\bibfnamefont {K.}~\bibnamefont {Wang}},\ and\ \bibinfo {author}
  {\bibfnamefont {J.}~\bibnamefont {Zhu}},\ }\bibfield  {title} {\bibinfo
  {title} {{Light dark matter confronted with the 95 GeV diphoton excess}},\
  }\href@noop {} {\  (\bibinfo {year} {2023})},\ \Eprint
  {https://arxiv.org/abs/2312.17599} {arXiv:2312.17599 [hep-ph]} \BibitemShut
  {NoStop}%
\bibitem [{\citenamefont {Ellwanger}\ \emph {et~al.}(2024)\citenamefont
  {Ellwanger}, \citenamefont {Hugonie}, \citenamefont {King},\ and\
  \citenamefont {Moretti}}]{Ellwanger:2024vvs}%
  \BibitemOpen
  \bibfield  {author} {\bibinfo {author} {\bibfnamefont {U.}~\bibnamefont
  {Ellwanger}}, \bibinfo {author} {\bibfnamefont {C.}~\bibnamefont {Hugonie}},
  \bibinfo {author} {\bibfnamefont {S.~F.}\ \bibnamefont {King}},\ and\
  \bibinfo {author} {\bibfnamefont {S.}~\bibnamefont {Moretti}},\ }\bibfield
  {title} {\bibinfo {title} {{NMSSM explanation for excesses in the search for
  neutralinos and charginos and a 95 GeV Higgs boson}},\ }\href
  {https://doi.org/10.1140/epjc/s10052-024-13129-2} {\bibfield  {journal}
  {\bibinfo  {journal} {Eur. Phys. J. C}\ }\textbf {\bibinfo {volume} {84}},\
  \bibinfo {pages} {788} (\bibinfo {year} {2024})},\ \Eprint
  {https://arxiv.org/abs/2404.19338} {arXiv:2404.19338 [hep-ph]} \BibitemShut
  {NoStop}%
\bibitem [{\citenamefont {Lian}(2024)}]{Lian:2024smg}%
  \BibitemOpen
  \bibfield  {author} {\bibinfo {author} {\bibfnamefont {J.}~\bibnamefont
  {Lian}},\ }\bibfield  {title} {\bibinfo {title} {{95~GeV excesses in the
  Z3-symmetric next-to-minimal supersymmetric standard model}},\ }\href
  {https://doi.org/10.1103/PhysRevD.110.115018} {\bibfield  {journal} {\bibinfo
   {journal} {Phys. Rev. D}\ }\textbf {\bibinfo {volume} {110}},\ \bibinfo
  {pages} {115018} (\bibinfo {year} {2024})},\ \Eprint
  {https://arxiv.org/abs/2406.10969} {arXiv:2406.10969 [hep-ph]} \BibitemShut
  {NoStop}%
\bibitem [{\citenamefont {Ellwanger}\ and\ \citenamefont
  {Hugonie}(2024)}]{Ellwanger:2024txc}%
  \BibitemOpen
  \bibfield  {author} {\bibinfo {author} {\bibfnamefont {U.}~\bibnamefont
  {Ellwanger}}\ and\ \bibinfo {author} {\bibfnamefont {C.}~\bibnamefont
  {Hugonie}},\ }\bibfield  {title} {\bibinfo {title} {{Nmssm with correct relic
  density and an additional 95~GeV Higgs boson}},\ }\href
  {https://doi.org/10.1140/epjc/s10052-024-12886-4} {\bibfield  {journal}
  {\bibinfo  {journal} {Eur. Phys. J. C}\ }\textbf {\bibinfo {volume} {84}},\
  \bibinfo {pages} {526} (\bibinfo {year} {2024})},\ \Eprint
  {https://arxiv.org/abs/2403.16884} {arXiv:2403.16884 [hep-ph]} \BibitemShut
  {NoStop}%
\bibitem [{\citenamefont {Cao}\ \emph {et~al.}(2024{\natexlab{a}})\citenamefont
  {Cao}, \citenamefont {Jia},\ and\ \citenamefont {Lian}}]{Cao:2024axg}%
  \BibitemOpen
  \bibfield  {author} {\bibinfo {author} {\bibfnamefont {J.}~\bibnamefont
  {Cao}}, \bibinfo {author} {\bibfnamefont {X.}~\bibnamefont {Jia}},\ and\
  \bibinfo {author} {\bibfnamefont {J.}~\bibnamefont {Lian}},\ }\bibfield
  {title} {\bibinfo {title} {{Unified interpretation of the muon g-2 anomaly,
  the 95~GeV diphoton, and bb\textasciimacron{} excesses in the general
  next-to-minimal supersymmetric standard model}},\ }\href
  {https://doi.org/10.1103/PhysRevD.110.115039} {\bibfield  {journal} {\bibinfo
   {journal} {Phys. Rev. D}\ }\textbf {\bibinfo {volume} {110}},\ \bibinfo
  {pages} {115039} (\bibinfo {year} {2024}{\natexlab{a}})},\ \Eprint
  {https://arxiv.org/abs/2402.15847} {arXiv:2402.15847 [hep-ph]} \BibitemShut
  {NoStop}%
\bibitem [{\citenamefont {Hammad}\ \emph {et~al.}(2025)\citenamefont {Hammad},
  \citenamefont {Ramos}, \citenamefont {Chakraborty}, \citenamefont {Ko},\ and\
  \citenamefont {Moretti}}]{Hammad:2025wst}%
  \BibitemOpen
  \bibfield  {author} {\bibinfo {author} {\bibfnamefont {A.}~\bibnamefont
  {Hammad}}, \bibinfo {author} {\bibfnamefont {R.}~\bibnamefont {Ramos}},
  \bibinfo {author} {\bibfnamefont {A.}~\bibnamefont {Chakraborty}}, \bibinfo
  {author} {\bibfnamefont {P.}~\bibnamefont {Ko}},\ and\ \bibinfo {author}
  {\bibfnamefont {S.}~\bibnamefont {Moretti}},\ }\bibfield  {title} {\bibinfo
  {title} {{Explaining Data Anomalies over the NMSSM Parameter Space with Deep
  Learning Techniques}},\ }\href@noop {} {\  (\bibinfo {year} {2025})},\
  \Eprint {https://arxiv.org/abs/2508.13912} {arXiv:2508.13912 [hep-ph]}
  \BibitemShut {NoStop}%
\bibitem [{\citenamefont {Cao}\ \emph {et~al.}(2024{\natexlab{b}})\citenamefont
  {Cao}, \citenamefont {Jia}, \citenamefont {Lian},\ and\ \citenamefont
  {Meng}}]{Cao:2023gkc}%
  \BibitemOpen
  \bibfield  {author} {\bibinfo {author} {\bibfnamefont {J.}~\bibnamefont
  {Cao}}, \bibinfo {author} {\bibfnamefont {X.}~\bibnamefont {Jia}}, \bibinfo
  {author} {\bibfnamefont {J.}~\bibnamefont {Lian}},\ and\ \bibinfo {author}
  {\bibfnamefont {L.}~\bibnamefont {Meng}},\ }\bibfield  {title} {\bibinfo
  {title} {{95~GeV diphoton and bb\textasciimacron{} excesses in the general
  next-to-minimal supersymmetric standard model}},\ }\href
  {https://doi.org/10.1103/PhysRevD.109.075001} {\bibfield  {journal} {\bibinfo
   {journal} {Phys. Rev. D}\ }\textbf {\bibinfo {volume} {109}},\ \bibinfo
  {pages} {075001} (\bibinfo {year} {2024}{\natexlab{b}})},\ \Eprint
  {https://arxiv.org/abs/2310.08436} {arXiv:2310.08436 [hep-ph]} \BibitemShut
  {NoStop}%
\bibitem [{\citenamefont {Ellwanger}\ and\ \citenamefont
  {Hugonie}(2023)}]{Ellwanger:2023zjc}%
  \BibitemOpen
  \bibfield  {author} {\bibinfo {author} {\bibfnamefont {U.}~\bibnamefont
  {Ellwanger}}\ and\ \bibinfo {author} {\bibfnamefont {C.}~\bibnamefont
  {Hugonie}},\ }\bibfield  {title} {\bibinfo {title} {{Additional Higgs Bosons
  near 95 and 650 GeV in the NMSSM}},\ }\href
  {https://doi.org/10.1140/epjc/s10052-023-12315-y} {\bibfield  {journal}
  {\bibinfo  {journal} {Eur. Phys. J. C}\ }\textbf {\bibinfo {volume} {83}},\
  \bibinfo {pages} {1138} (\bibinfo {year} {2023})},\ \Eprint
  {https://arxiv.org/abs/2309.07838} {arXiv:2309.07838 [hep-ph]} \BibitemShut
  {NoStop}%
\bibitem [{\citenamefont {Benbrik}\ \emph {et~al.}(2025)\citenamefont
  {Benbrik}, \citenamefont {Boukidi}, \citenamefont {Kahime}, \citenamefont
  {Moretti}, \citenamefont {Rahili},\ and\ \citenamefont
  {Taki}}]{Benbrik:2025hol}%
  \BibitemOpen
  \bibfield  {author} {\bibinfo {author} {\bibfnamefont {R.}~\bibnamefont
  {Benbrik}}, \bibinfo {author} {\bibfnamefont {M.}~\bibnamefont {Boukidi}},
  \bibinfo {author} {\bibfnamefont {K.}~\bibnamefont {Kahime}}, \bibinfo
  {author} {\bibfnamefont {S.}~\bibnamefont {Moretti}}, \bibinfo {author}
  {\bibfnamefont {L.}~\bibnamefont {Rahili}},\ and\ \bibinfo {author}
  {\bibfnamefont {B.}~\bibnamefont {Taki}},\ }\bibfield  {title} {\bibinfo
  {title} {{Exploring potential Higgs resonances at 650 GeV and 95 GeV in the
  2HDM Type III}},\ }\href {https://doi.org/10.1016/j.physletb.2025.139688}
  {\bibfield  {journal} {\bibinfo  {journal} {Phys. Lett. B}\ }\textbf
  {\bibinfo {volume} {868}},\ \bibinfo {pages} {139688} (\bibinfo {year}
  {2025})},\ \Eprint {https://arxiv.org/abs/2505.07811} {arXiv:2505.07811
  [hep-ph]} \BibitemShut {NoStop}%
\bibitem [{\citenamefont {Tumasyan}\ \emph {et~al.}(2021)\citenamefont
  {Tumasyan} \emph {et~al.}}]{CMS:2021yci}%
  \BibitemOpen
  \bibfield  {author} {\bibinfo {author} {\bibfnamefont {A.}~\bibnamefont
  {Tumasyan}} \emph {et~al.} (\bibinfo {collaboration} {CMS}),\ }\bibfield
  {title} {\bibinfo {title} {{Search for a heavy Higgs boson decaying into two
  lighter Higgs bosons in the $\tau\tau$bb final state at 13 TeV}},\ }\href
  {https://doi.org/10.1007/JHEP11(2021)057} {\bibfield  {journal} {\bibinfo
  {journal} {JHEP}\ }\textbf {\bibinfo {volume} {11}},\ \bibinfo {pages}
  {057}},\ \Eprint {https://arxiv.org/abs/2106.10361} {arXiv:2106.10361
  [hep-ex]} \BibitemShut {NoStop}%
\bibitem [{\citenamefont {Hayrapetyan}\ \emph
  {et~al.}(2025{\natexlab{b}})\citenamefont {Hayrapetyan} \emph
  {et~al.}}]{CMS:2025tqi}%
  \BibitemOpen
  \bibfield  {author} {\bibinfo {author} {\bibfnamefont {A.}~\bibnamefont
  {Hayrapetyan}} \emph {et~al.} (\bibinfo {collaboration} {CMS}),\ }\bibfield
  {title} {\bibinfo {title} {{Search for the nonresonant and resonant
  production of a Higgs boson in association with an additional scalar boson in
  the $\gamma \gamma \tau \tau$ final state in proton-proton collisions at
  $\sqrt{s}$ = 13 TeV}},\ }\href@noop {} {\  (\bibinfo {year}
  {2025}{\natexlab{b}})},\ \Eprint {https://arxiv.org/abs/2506.23012}
  {arXiv:2506.23012 [hep-ex]} \BibitemShut {NoStop}%
\bibitem [{\citenamefont {Hayrapetyan}\ \emph
  {et~al.}(2025{\natexlab{c}})\citenamefont {Hayrapetyan} \emph
  {et~al.}}]{CMS:2025qit}%
  \BibitemOpen
  \bibfield  {author} {\bibinfo {author} {\bibfnamefont {A.}~\bibnamefont
  {Hayrapetyan}} \emph {et~al.} (\bibinfo {collaboration} {CMS}),\ }\bibfield
  {title} {\bibinfo {title} {{Search for a new scalar resonance decaying to a
  Higgs boson and another new scalar particle in the final state with two
  bottom quarks and two photons in proton-proton collisions at $\sqrt{s}$ = 13
  TeV}},\ }\href@noop {} {\  (\bibinfo {year} {2025}{\natexlab{c}})},\ \Eprint
  {https://arxiv.org/abs/2508.11494} {arXiv:2508.11494 [hep-ex]} \BibitemShut
  {NoStop}%
\bibitem [{Note2()}]{Note2}%
  \BibitemOpen
  \bibinfo {note} {And the I(1+2)HDM used here is CP-conserving, so that the
  decay $A\to h_{125}h_{95}$ is not possible.}\BibitemShut {Stop}%
\bibitem [{\citenamefont {Landau}(1948)}]{Landau:1948kw}%
  \BibitemOpen
  \bibfield  {author} {\bibinfo {author} {\bibfnamefont {L.~D.}\ \bibnamefont
  {Landau}},\ }\bibfield  {title} {\bibinfo {title} {{On the angular momentum
  of a system of two photons}},\ }\href
  {https://doi.org/10.1016/B978-0-08-010586-4.50070-5} {\bibfield  {journal}
  {\bibinfo  {journal} {Dokl. Akad. Nauk SSSR}\ }\textbf {\bibinfo {volume}
  {60}},\ \bibinfo {pages} {207} (\bibinfo {year} {1948})}\BibitemShut
  {NoStop}%
\bibitem [{\citenamefont {Yang}(1950)}]{Yang:1950rg}%
  \BibitemOpen
  \bibfield  {author} {\bibinfo {author} {\bibfnamefont {C.-N.}\ \bibnamefont
  {Yang}},\ }\bibfield  {title} {\bibinfo {title} {{Selection Rules for the
  Dematerialization of a Particle Into Two Photons}},\ }\href
  {https://doi.org/10.1103/PhysRev.77.242} {\bibfield  {journal} {\bibinfo
  {journal} {Phys. Rev.}\ }\textbf {\bibinfo {volume} {77}},\ \bibinfo {pages}
  {242} (\bibinfo {year} {1950})}\BibitemShut {NoStop}%
\bibitem [{\citenamefont {Moretti}(2015)}]{Moretti:2014rka}%
  \BibitemOpen
  \bibfield  {author} {\bibinfo {author} {\bibfnamefont {S.}~\bibnamefont
  {Moretti}},\ }\bibfield  {title} {\bibinfo {title} {{Variations on a Higgs
  theme}},\ }\href {https://doi.org/10.1103/PhysRevD.91.014012} {\bibfield
  {journal} {\bibinfo  {journal} {Phys. Rev. D}\ }\textbf {\bibinfo {volume}
  {91}},\ \bibinfo {pages} {014012} (\bibinfo {year} {2015})},\ \Eprint
  {https://arxiv.org/abs/1407.3511} {arXiv:1407.3511 [hep-ph]} \BibitemShut
  {NoStop}%
\bibitem [{\citenamefont {Branco}\ \emph {et~al.}(2012)\citenamefont {Branco},
  \citenamefont {Ferreira}, \citenamefont {Lavoura}, \citenamefont {Rebelo},
  \citenamefont {Sher},\ and\ \citenamefont {Silva}}]{Branco:2011iw}%
  \BibitemOpen
  \bibfield  {author} {\bibinfo {author} {\bibfnamefont {G.~C.}\ \bibnamefont
  {Branco}}, \bibinfo {author} {\bibfnamefont {P.~M.}\ \bibnamefont
  {Ferreira}}, \bibinfo {author} {\bibfnamefont {L.}~\bibnamefont {Lavoura}},
  \bibinfo {author} {\bibfnamefont {M.~N.}\ \bibnamefont {Rebelo}}, \bibinfo
  {author} {\bibfnamefont {M.}~\bibnamefont {Sher}},\ and\ \bibinfo {author}
  {\bibfnamefont {J.~P.}\ \bibnamefont {Silva}},\ }\bibfield  {title} {\bibinfo
  {title} {{Theory and phenomenology of two-Higgs-doublet models}},\ }\href
  {https://doi.org/10.1016/j.physrep.2012.02.002} {\bibfield  {journal}
  {\bibinfo  {journal} {Phys. Rept.}\ }\textbf {\bibinfo {volume} {516}},\
  \bibinfo {pages} {1} (\bibinfo {year} {2012})},\ \Eprint
  {https://arxiv.org/abs/1106.0034} {arXiv:1106.0034 [hep-ph]} \BibitemShut
  {NoStop}%
\bibitem [{\citenamefont {Grzadkowski}\ \emph {et~al.}(2009)\citenamefont
  {Grzadkowski}, \citenamefont {Ogreid},\ and\ \citenamefont
  {Osland}}]{Grzadkowski:2009bt}%
  \BibitemOpen
  \bibfield  {author} {\bibinfo {author} {\bibfnamefont {B.}~\bibnamefont
  {Grzadkowski}}, \bibinfo {author} {\bibfnamefont {O.~M.}\ \bibnamefont
  {Ogreid}},\ and\ \bibinfo {author} {\bibfnamefont {P.}~\bibnamefont
  {Osland}},\ }\bibfield  {title} {\bibinfo {title} {{Natural Multi-Higgs Model
  with Dark Matter and CP Violation}},\ }\href
  {https://doi.org/10.1103/PhysRevD.80.055013} {\bibfield  {journal} {\bibinfo
  {journal} {Phys. Rev. D}\ }\textbf {\bibinfo {volume} {80}},\ \bibinfo
  {pages} {055013} (\bibinfo {year} {2009})},\ \Eprint
  {https://arxiv.org/abs/0904.2173} {arXiv:0904.2173 [hep-ph]} \BibitemShut
  {NoStop}%
\bibitem [{\citenamefont {Moretti}\ and\ \citenamefont
  {Yagyu}(2015)}]{Moretti:2015cwa}%
  \BibitemOpen
  \bibfield  {author} {\bibinfo {author} {\bibfnamefont {S.}~\bibnamefont
  {Moretti}}\ and\ \bibinfo {author} {\bibfnamefont {K.}~\bibnamefont
  {Yagyu}},\ }\bibfield  {title} {\bibinfo {title} {{Constraints on Parameter
  Space from Perturbative Unitarity in Models with Three Scalar Doublets}},\
  }\href {https://doi.org/10.1103/PhysRevD.91.055022} {\bibfield  {journal}
  {\bibinfo  {journal} {Phys. Rev. D}\ }\textbf {\bibinfo {volume} {91}},\
  \bibinfo {pages} {055022} (\bibinfo {year} {2015})},\ \Eprint
  {https://arxiv.org/abs/1501.06544} {arXiv:1501.06544 [hep-ph]} \BibitemShut
  {NoStop}%
\bibitem [{\citenamefont {Glashow}\ and\ \citenamefont
  {Weinberg}(1977)}]{Glashow:1976nt}%
  \BibitemOpen
  \bibfield  {author} {\bibinfo {author} {\bibfnamefont {S.~L.}\ \bibnamefont
  {Glashow}}\ and\ \bibinfo {author} {\bibfnamefont {S.}~\bibnamefont
  {Weinberg}},\ }\bibfield  {title} {\bibinfo {title} {{Natural Conservation
  Laws for Neutral Currents}},\ }\href
  {https://doi.org/10.1103/PhysRevD.15.1958} {\bibfield  {journal} {\bibinfo
  {journal} {Phys. Rev. D}\ }\textbf {\bibinfo {volume} {15}},\ \bibinfo
  {pages} {1958} (\bibinfo {year} {1977})}\BibitemShut {NoStop}%
\bibitem [{\citenamefont {Gunion}\ and\ \citenamefont
  {Haber}(2003)}]{Gunion:2002zf}%
  \BibitemOpen
  \bibfield  {author} {\bibinfo {author} {\bibfnamefont {J.~F.}\ \bibnamefont
  {Gunion}}\ and\ \bibinfo {author} {\bibfnamefont {H.~E.}\ \bibnamefont
  {Haber}},\ }\bibfield  {title} {\bibinfo {title} {{The CP conserving two
  Higgs doublet model: The Approach to the decoupling limit}},\ }\href
  {https://doi.org/10.1103/PhysRevD.67.075019} {\bibfield  {journal} {\bibinfo
  {journal} {Phys. Rev. D}\ }\textbf {\bibinfo {volume} {67}},\ \bibinfo
  {pages} {075019} (\bibinfo {year} {2003})},\ \Eprint
  {https://arxiv.org/abs/hep-ph/0207010} {arXiv:hep-ph/0207010} \BibitemShut
  {NoStop}%
\bibitem [{\citenamefont {Keus}\ \emph {et~al.}(2014)\citenamefont {Keus},
  \citenamefont {King}, \citenamefont {Moretti},\ and\ \citenamefont
  {Sokolowska}}]{Keus:2014jha}%
  \BibitemOpen
  \bibfield  {author} {\bibinfo {author} {\bibfnamefont {V.}~\bibnamefont
  {Keus}}, \bibinfo {author} {\bibfnamefont {S.~F.}\ \bibnamefont {King}},
  \bibinfo {author} {\bibfnamefont {S.}~\bibnamefont {Moretti}},\ and\ \bibinfo
  {author} {\bibfnamefont {D.}~\bibnamefont {Sokolowska}},\ }\bibfield  {title}
  {\bibinfo {title} {{Dark Matter with Two Inert Doublets plus One Higgs
  Doublet}},\ }\href {https://doi.org/10.1007/JHEP11(2014)016} {\bibfield
  {journal} {\bibinfo  {journal} {JHEP}\ }\textbf {\bibinfo {volume} {11}},\
  \bibinfo {pages} {016}},\ \Eprint {https://arxiv.org/abs/1407.7859}
  {arXiv:1407.7859 [hep-ph]} \BibitemShut {NoStop}%
\bibitem [{\citenamefont {Keus}\ \emph {et~al.}(2015)\citenamefont {Keus},
  \citenamefont {King}, \citenamefont {Moretti},\ and\ \citenamefont
  {Sokolowska}}]{Keus:2015xya}%
  \BibitemOpen
  \bibfield  {author} {\bibinfo {author} {\bibfnamefont {V.}~\bibnamefont
  {Keus}}, \bibinfo {author} {\bibfnamefont {S.~F.}\ \bibnamefont {King}},
  \bibinfo {author} {\bibfnamefont {S.}~\bibnamefont {Moretti}},\ and\ \bibinfo
  {author} {\bibfnamefont {D.}~\bibnamefont {Sokolowska}},\ }\bibfield  {title}
  {\bibinfo {title} {{Observable Heavy Higgs Dark Matter}},\ }\href
  {https://doi.org/10.1007/JHEP11(2015)003} {\bibfield  {journal} {\bibinfo
  {journal} {JHEP}\ }\textbf {\bibinfo {volume} {11}},\ \bibinfo {pages}
  {003}},\ \Eprint {https://arxiv.org/abs/1507.08433} {arXiv:1507.08433
  [hep-ph]} \BibitemShut {NoStop}%
\bibitem [{\citenamefont {Cordero-Cid}\ \emph {et~al.}(2016)\citenamefont
  {Cordero-Cid}, \citenamefont {Hern{\'a}ndez-S{\'a}nchez}, \citenamefont
  {Keus}, \citenamefont {King}, \citenamefont {Moretti}, \citenamefont
  {Rojas},\ and\ \citenamefont {Soko{\l}owska}}]{Cordero-Cid:2016krd}%
  \BibitemOpen
  \bibfield  {author} {\bibinfo {author} {\bibfnamefont {A.}~\bibnamefont
  {Cordero-Cid}}, \bibinfo {author} {\bibfnamefont {J.}~\bibnamefont
  {Hern{\'a}ndez-S{\'a}nchez}}, \bibinfo {author} {\bibfnamefont
  {V.}~\bibnamefont {Keus}}, \bibinfo {author} {\bibfnamefont {S.~F.}\
  \bibnamefont {King}}, \bibinfo {author} {\bibfnamefont {S.}~\bibnamefont
  {Moretti}}, \bibinfo {author} {\bibfnamefont {D.}~\bibnamefont {Rojas}},\
  and\ \bibinfo {author} {\bibfnamefont {D.}~\bibnamefont {Soko{\l}owska}},\
  }\bibfield  {title} {\bibinfo {title} {{CP violating scalar Dark Matter}},\
  }\href {https://doi.org/10.1007/JHEP12(2016)014} {\bibfield  {journal}
  {\bibinfo  {journal} {JHEP}\ }\textbf {\bibinfo {volume} {12}},\ \bibinfo
  {pages} {014}},\ \Eprint {https://arxiv.org/abs/1608.01673} {arXiv:1608.01673
  [hep-ph]} \BibitemShut {NoStop}%
\bibitem [{\citenamefont {Cordero}\ \emph {et~al.}(2018)\citenamefont
  {Cordero}, \citenamefont {Hernandez-Sanchez}, \citenamefont {Keus},
  \citenamefont {King}, \citenamefont {Moretti}, \citenamefont {Rojas},\ and\
  \citenamefont {Sokolowska}}]{Cordero:2017owj}%
  \BibitemOpen
  \bibfield  {author} {\bibinfo {author} {\bibfnamefont {A.}~\bibnamefont
  {Cordero}}, \bibinfo {author} {\bibfnamefont {J.}~\bibnamefont
  {Hernandez-Sanchez}}, \bibinfo {author} {\bibfnamefont {V.}~\bibnamefont
  {Keus}}, \bibinfo {author} {\bibfnamefont {S.~F.}\ \bibnamefont {King}},
  \bibinfo {author} {\bibfnamefont {S.}~\bibnamefont {Moretti}}, \bibinfo
  {author} {\bibfnamefont {D.}~\bibnamefont {Rojas}},\ and\ \bibinfo {author}
  {\bibfnamefont {D.}~\bibnamefont {Sokolowska}},\ }\bibfield  {title}
  {\bibinfo {title} {{Dark Matter Signals at the LHC from a 3HDM}},\ }\href
  {https://doi.org/10.1007/JHEP05(2018)030} {\bibfield  {journal} {\bibinfo
  {journal} {JHEP}\ }\textbf {\bibinfo {volume} {05}},\ \bibinfo {pages}
  {030}},\ \Eprint {https://arxiv.org/abs/1712.09598} {arXiv:1712.09598
  [hep-ph]} \BibitemShut {NoStop}%
\bibitem [{\citenamefont {Bechtle}\ \emph {et~al.}(2021)\citenamefont
  {Bechtle}, \citenamefont {Heinemeyer}, \citenamefont {Klingl}, \citenamefont
  {Stefaniak}, \citenamefont {Weiglein},\ and\ \citenamefont
  {Wittbrodt}}]{Bechtle:2020uwn}%
  \BibitemOpen
  \bibfield  {author} {\bibinfo {author} {\bibfnamefont {P.}~\bibnamefont
  {Bechtle}}, \bibinfo {author} {\bibfnamefont {S.}~\bibnamefont {Heinemeyer}},
  \bibinfo {author} {\bibfnamefont {T.}~\bibnamefont {Klingl}}, \bibinfo
  {author} {\bibfnamefont {T.}~\bibnamefont {Stefaniak}}, \bibinfo {author}
  {\bibfnamefont {G.}~\bibnamefont {Weiglein}},\ and\ \bibinfo {author}
  {\bibfnamefont {J.}~\bibnamefont {Wittbrodt}},\ }\bibfield  {title} {\bibinfo
  {title} {{HiggsSignals-2: Probing new physics with precision Higgs
  measurements in the LHC 13 TeV era}},\ }\href
  {https://doi.org/10.1140/epjc/s10052-021-08942-y} {\bibfield  {journal}
  {\bibinfo  {journal} {Eur. Phys. J. C}\ }\textbf {\bibinfo {volume} {81}},\
  \bibinfo {pages} {145} (\bibinfo {year} {2021})},\ \Eprint
  {https://arxiv.org/abs/2012.09197} {arXiv:2012.09197 [hep-ph]} \BibitemShut
  {NoStop}%
\bibitem [{\citenamefont {Bechtle}\ \emph {et~al.}(2020)\citenamefont
  {Bechtle}, \citenamefont {Dercks}, \citenamefont {Heinemeyer}, \citenamefont
  {Klingl}, \citenamefont {Stefaniak}, \citenamefont {Weiglein},\ and\
  \citenamefont {Wittbrodt}}]{Bechtle:2020pkv}%
  \BibitemOpen
  \bibfield  {author} {\bibinfo {author} {\bibfnamefont {P.}~\bibnamefont
  {Bechtle}}, \bibinfo {author} {\bibfnamefont {D.}~\bibnamefont {Dercks}},
  \bibinfo {author} {\bibfnamefont {S.}~\bibnamefont {Heinemeyer}}, \bibinfo
  {author} {\bibfnamefont {T.}~\bibnamefont {Klingl}}, \bibinfo {author}
  {\bibfnamefont {T.}~\bibnamefont {Stefaniak}}, \bibinfo {author}
  {\bibfnamefont {G.}~\bibnamefont {Weiglein}},\ and\ \bibinfo {author}
  {\bibfnamefont {J.}~\bibnamefont {Wittbrodt}},\ }\bibfield  {title} {\bibinfo
  {title} {{HiggsBounds-5: Testing Higgs Sectors in the LHC 13 TeV Era}},\
  }\href {https://doi.org/10.1140/epjc/s10052-020-08557-9} {\bibfield
  {journal} {\bibinfo  {journal} {Eur. Phys. J. C}\ }\textbf {\bibinfo {volume}
  {80}},\ \bibinfo {pages} {1211} (\bibinfo {year} {2020})},\ \Eprint
  {https://arxiv.org/abs/2006.06007} {arXiv:2006.06007 [hep-ph]} \BibitemShut
  {NoStop}%
\bibitem [{\citenamefont {Bahl}\ \emph {et~al.}(2023)\citenamefont {Bahl},
  \citenamefont {Biek\"otter}, \citenamefont {Heinemeyer}, \citenamefont {Li},
  \citenamefont {Paasch}, \citenamefont {Weiglein},\ and\ \citenamefont
  {Wittbrodt}}]{Bahl:2022igd}%
  \BibitemOpen
  \bibfield  {author} {\bibinfo {author} {\bibfnamefont {H.}~\bibnamefont
  {Bahl}}, \bibinfo {author} {\bibfnamefont {T.}~\bibnamefont {Biek\"otter}},
  \bibinfo {author} {\bibfnamefont {S.}~\bibnamefont {Heinemeyer}}, \bibinfo
  {author} {\bibfnamefont {C.}~\bibnamefont {Li}}, \bibinfo {author}
  {\bibfnamefont {S.}~\bibnamefont {Paasch}}, \bibinfo {author} {\bibfnamefont
  {G.}~\bibnamefont {Weiglein}},\ and\ \bibinfo {author} {\bibfnamefont
  {J.}~\bibnamefont {Wittbrodt}},\ }\bibfield  {title} {\bibinfo {title}
  {{HiggsTools: BSM scalar phenomenology with new versions of HiggsBounds and
  HiggsSignals}},\ }\href {https://doi.org/10.1016/j.cpc.2023.108803}
  {\bibfield  {journal} {\bibinfo  {journal} {Comput. Phys. Commun.}\ }\textbf
  {\bibinfo {volume} {291}},\ \bibinfo {pages} {108803} (\bibinfo {year}
  {2023})},\ \Eprint {https://arxiv.org/abs/2210.09332} {arXiv:2210.09332
  [hep-ph]} \BibitemShut {NoStop}%
\bibitem [{\citenamefont {Merchand}\ and\ \citenamefont
  {Sher}(2020)}]{Merchand:2019bod}%
  \BibitemOpen
  \bibfield  {author} {\bibinfo {author} {\bibfnamefont {M.}~\bibnamefont
  {Merchand}}\ and\ \bibinfo {author} {\bibfnamefont {M.}~\bibnamefont
  {Sher}},\ }\bibfield  {title} {\bibinfo {title} {{Constraints on the
  Parameter Space in an Inert Doublet Model with two Active Doublets}},\ }\href
  {https://doi.org/10.1007/JHEP03(2020)108} {\bibfield  {journal} {\bibinfo
  {journal} {JHEP}\ }\textbf {\bibinfo {volume} {03}},\ \bibinfo {pages}
  {108}},\ \Eprint {https://arxiv.org/abs/1911.06477} {arXiv:1911.06477
  [hep-ph]} \BibitemShut {NoStop}%
\bibitem [{\citenamefont {Peskin}\ and\ \citenamefont
  {Takeuchi}(1992)}]{Peskin:1991sw}%
  \BibitemOpen
  \bibfield  {author} {\bibinfo {author} {\bibfnamefont {M.~E.}\ \bibnamefont
  {Peskin}}\ and\ \bibinfo {author} {\bibfnamefont {T.}~\bibnamefont
  {Takeuchi}},\ }\bibfield  {title} {\bibinfo {title} {{Estimation of oblique
  electroweak corrections}},\ }\href {https://doi.org/10.1103/PhysRevD.46.381}
  {\bibfield  {journal} {\bibinfo  {journal} {Phys. Rev. D}\ }\textbf {\bibinfo
  {volume} {46}},\ \bibinfo {pages} {381} (\bibinfo {year} {1992})}\BibitemShut
  {NoStop}%
\bibitem [{\citenamefont {Grimus}\ \emph {et~al.}(2008)\citenamefont {Grimus},
  \citenamefont {Lavoura}, \citenamefont {Ogreid},\ and\ \citenamefont
  {Osland}}]{Grimus:2008nb}%
  \BibitemOpen
  \bibfield  {author} {\bibinfo {author} {\bibfnamefont {W.}~\bibnamefont
  {Grimus}}, \bibinfo {author} {\bibfnamefont {L.}~\bibnamefont {Lavoura}},
  \bibinfo {author} {\bibfnamefont {O.~M.}\ \bibnamefont {Ogreid}},\ and\
  \bibinfo {author} {\bibfnamefont {P.}~\bibnamefont {Osland}},\ }\bibfield
  {title} {\bibinfo {title} {{The Oblique parameters in multi-Higgs-doublet
  models}},\ }\href {https://doi.org/10.1016/j.nuclphysb.2008.04.019}
  {\bibfield  {journal} {\bibinfo  {journal} {Nucl. Phys. B}\ }\textbf
  {\bibinfo {volume} {801}},\ \bibinfo {pages} {81} (\bibinfo {year} {2008})},\
  \Eprint {https://arxiv.org/abs/0802.4353} {arXiv:0802.4353 [hep-ph]}
  \BibitemShut {NoStop}%
\bibitem [{\citenamefont {Alguero}\ \emph {et~al.}(2024)\citenamefont
  {Alguero}, \citenamefont {Belanger}, \citenamefont {Boudjema}, \citenamefont
  {Chakraborti}, \citenamefont {Goudelis}, \citenamefont {Kraml}, \citenamefont
  {Mjallal},\ and\ \citenamefont {Pukhov}}]{Alguero:2023zol}%
  \BibitemOpen
  \bibfield  {author} {\bibinfo {author} {\bibfnamefont {G.}~\bibnamefont
  {Alguero}}, \bibinfo {author} {\bibfnamefont {G.}~\bibnamefont {Belanger}},
  \bibinfo {author} {\bibfnamefont {F.}~\bibnamefont {Boudjema}}, \bibinfo
  {author} {\bibfnamefont {S.}~\bibnamefont {Chakraborti}}, \bibinfo {author}
  {\bibfnamefont {A.}~\bibnamefont {Goudelis}}, \bibinfo {author}
  {\bibfnamefont {S.}~\bibnamefont {Kraml}}, \bibinfo {author} {\bibfnamefont
  {A.}~\bibnamefont {Mjallal}},\ and\ \bibinfo {author} {\bibfnamefont
  {A.}~\bibnamefont {Pukhov}},\ }\bibfield  {title} {\bibinfo {title}
  {{micrOMEGAs 6.0: N-component dark matter}},\ }\href
  {https://doi.org/10.1016/j.cpc.2024.109133} {\bibfield  {journal} {\bibinfo
  {journal} {Comput. Phys. Commun.}\ }\textbf {\bibinfo {volume} {299}},\
  \bibinfo {pages} {109133} (\bibinfo {year} {2024})},\ \Eprint
  {https://arxiv.org/abs/2312.14894} {arXiv:2312.14894 [hep-ph]} \BibitemShut
  {NoStop}%
\bibitem [{\citenamefont {Mahmoudi}(2009)}]{Mahmoudi:2008tp}%
  \BibitemOpen
  \bibfield  {author} {\bibinfo {author} {\bibfnamefont {F.}~\bibnamefont
  {Mahmoudi}},\ }\bibfield  {title} {\bibinfo {title} {{SuperIso v2.3: A
  Program for calculating flavor physics observables in Supersymmetry}},\
  }\href {https://doi.org/10.1016/j.cpc.2009.02.017} {\bibfield  {journal}
  {\bibinfo  {journal} {Comput. Phys. Commun.}\ }\textbf {\bibinfo {volume}
  {180}},\ \bibinfo {pages} {1579} (\bibinfo {year} {2009})},\ \Eprint
  {https://arxiv.org/abs/0808.3144} {arXiv:0808.3144 [hep-ph]} \BibitemShut
  {NoStop}%
\bibitem [{\citenamefont {Amhis}\ \emph {et~al.}(2017)\citenamefont {Amhis}
  \emph {et~al.}}]{HFLAV:2016hnz}%
  \BibitemOpen
  \bibfield  {author} {\bibinfo {author} {\bibfnamefont {Y.}~\bibnamefont
  {Amhis}} \emph {et~al.} (\bibinfo {collaboration} {HFLAV}),\ }\bibfield
  {title} {\bibinfo {title} {{Averages of $b$-hadron, $c$-hadron, and
  $\tau$-lepton properties as of summer 2016}},\ }\href
  {https://doi.org/10.1140/epjc/s10052-017-5058-4} {\bibfield  {journal}
  {\bibinfo  {journal} {Eur. Phys. J. C}\ }\textbf {\bibinfo {volume} {77}},\
  \bibinfo {pages} {895} (\bibinfo {year} {2017})},\ \Eprint
  {https://arxiv.org/abs/1612.07233} {arXiv:1612.07233 [hep-ex]} \BibitemShut
  {NoStop}%
\bibitem [{\citenamefont {Aaij}\ \emph
  {et~al.}(2022{\natexlab{a}})\citenamefont {Aaij} \emph
  {et~al.}}]{LHCb:2021awg}%
  \BibitemOpen
  \bibfield  {author} {\bibinfo {author} {\bibfnamefont {R.}~\bibnamefont
  {Aaij}} \emph {et~al.} (\bibinfo {collaboration} {LHCb}),\ }\bibfield
  {title} {\bibinfo {title} {{Measurement of the $B^0_s\to\mu^+\mu^-$ decay
  properties and search for the $B^0\to\mu^+\mu^-$ and
  $B^0_s\to\mu^+\mu^-\gamma$ decays}},\ }\href
  {https://doi.org/10.1103/PhysRevD.105.012010} {\bibfield  {journal} {\bibinfo
   {journal} {Phys. Rev. D}\ }\textbf {\bibinfo {volume} {105}},\ \bibinfo
  {pages} {012010} (\bibinfo {year} {2022}{\natexlab{a}})},\ \Eprint
  {https://arxiv.org/abs/2108.09283} {arXiv:2108.09283 [hep-ex]} \BibitemShut
  {NoStop}%
\bibitem [{\citenamefont {Aaij}\ \emph
  {et~al.}(2022{\natexlab{b}})\citenamefont {Aaij} \emph
  {et~al.}}]{LHCb:2021vsc}%
  \BibitemOpen
  \bibfield  {author} {\bibinfo {author} {\bibfnamefont {R.}~\bibnamefont
  {Aaij}} \emph {et~al.} (\bibinfo {collaboration} {LHCb}),\ }\bibfield
  {title} {\bibinfo {title} {{Analysis of Neutral B-Meson Decays into Two
  Muons}},\ }\href {https://doi.org/10.1103/PhysRevLett.128.041801} {\bibfield
  {journal} {\bibinfo  {journal} {Phys. Rev. Lett.}\ }\textbf {\bibinfo
  {volume} {128}},\ \bibinfo {pages} {041801} (\bibinfo {year}
  {2022}{\natexlab{b}})},\ \Eprint {https://arxiv.org/abs/2108.09284}
  {arXiv:2108.09284 [hep-ex]} \BibitemShut {NoStop}%
\bibitem [{\citenamefont {Tumasyan}\ \emph
  {et~al.}(2023{\natexlab{b}})\citenamefont {Tumasyan} \emph
  {et~al.}}]{CMS:2022mgd}%
  \BibitemOpen
  \bibfield  {author} {\bibinfo {author} {\bibfnamefont {A.}~\bibnamefont
  {Tumasyan}} \emph {et~al.} (\bibinfo {collaboration} {CMS}),\ }\bibfield
  {title} {\bibinfo {title} {{Measurement of the $B^0_{s} \to \mu^+\mu^-$ decay
  properties and search for the $B^0 \to \mu^+\mu^-$ decay in proton-proton
  collisions at $\sqrt{s}$ = 13 TeV}},\ }\href
  {https://doi.org/10.1016/j.physletb.2023.137955} {\bibfield  {journal}
  {\bibinfo  {journal} {Phys. Lett. B}\ }\textbf {\bibinfo {volume} {842}},\
  \bibinfo {pages} {137955} (\bibinfo {year} {2023}{\natexlab{b}})},\ \Eprint
  {https://arxiv.org/abs/2212.10311} {arXiv:2212.10311 [hep-ex]} \BibitemShut
  {NoStop}%
\bibitem [{\citenamefont {Aprile}\ \emph {et~al.}(2023)\citenamefont {Aprile}
  \emph {et~al.}}]{XENON:2023cxc}%
  \BibitemOpen
  \bibfield  {author} {\bibinfo {author} {\bibfnamefont {E.}~\bibnamefont
  {Aprile}} \emph {et~al.} (\bibinfo {collaboration} {XENON}),\ }\bibfield
  {title} {\bibinfo {title} {{First Dark Matter Search with Nuclear Recoils
  from the XENONnT Experiment}},\ }\href
  {https://doi.org/10.1103/PhysRevLett.131.041003} {\bibfield  {journal}
  {\bibinfo  {journal} {Phys. Rev. Lett.}\ }\textbf {\bibinfo {volume} {131}},\
  \bibinfo {pages} {041003} (\bibinfo {year} {2023})},\ \Eprint
  {https://arxiv.org/abs/2303.14729} {arXiv:2303.14729 [hep-ex]} \BibitemShut
  {NoStop}%
\bibitem [{\citenamefont {Aalbers}\ \emph {et~al.}(2023)\citenamefont {Aalbers}
  \emph {et~al.}}]{LZ:2022lsv}%
  \BibitemOpen
  \bibfield  {author} {\bibinfo {author} {\bibfnamefont {J.}~\bibnamefont
  {Aalbers}} \emph {et~al.} (\bibinfo {collaboration} {LZ}),\ }\bibfield
  {title} {\bibinfo {title} {{First Dark Matter Search Results from the
  LUX-ZEPLIN (LZ) Experiment}},\ }\href
  {https://doi.org/10.1103/PhysRevLett.131.041002} {\bibfield  {journal}
  {\bibinfo  {journal} {Phys. Rev. Lett.}\ }\textbf {\bibinfo {volume} {131}},\
  \bibinfo {pages} {041002} (\bibinfo {year} {2023})},\ \Eprint
  {https://arxiv.org/abs/2207.03764} {arXiv:2207.03764 [hep-ex]} \BibitemShut
  {NoStop}%
\bibitem [{\citenamefont {Aalbers}\ \emph {et~al.}(2025)\citenamefont {Aalbers}
  \emph {et~al.}}]{LZ:2024zvo}%
  \BibitemOpen
  \bibfield  {author} {\bibinfo {author} {\bibfnamefont {J.}~\bibnamefont
  {Aalbers}} \emph {et~al.} (\bibinfo {collaboration} {LZ}),\ }\bibfield
  {title} {\bibinfo {title} {{Dark Matter Search Results from
  4.2{\,}{\,}Tonne-Years of Exposure of the LUX-ZEPLIN (LZ) Experiment}},\
  }\href {https://doi.org/10.1103/4dyc-z8zf} {\bibfield  {journal} {\bibinfo
  {journal} {Phys. Rev. Lett.}\ }\textbf {\bibinfo {volume} {135}},\ \bibinfo
  {pages} {011802} (\bibinfo {year} {2025})},\ \Eprint
  {https://arxiv.org/abs/2410.17036} {arXiv:2410.17036 [hep-ex]} \BibitemShut
  {NoStop}%
\bibitem [{\citenamefont {Aghanim}\ \emph {et~al.}(2020)\citenamefont {Aghanim}
  \emph {et~al.}}]{Planck:2018vyg}%
  \BibitemOpen
  \bibfield  {author} {\bibinfo {author} {\bibfnamefont {N.}~\bibnamefont
  {Aghanim}} \emph {et~al.} (\bibinfo {collaboration} {Planck}),\ }\bibfield
  {title} {\bibinfo {title} {{Planck 2018 results. VI. Cosmological
  parameters}},\ }\href {https://doi.org/10.1051/0004-6361/201833910}
  {\bibfield  {journal} {\bibinfo  {journal} {Astron. Astrophys.}\ }\textbf
  {\bibinfo {volume} {641}},\ \bibinfo {pages} {A6} (\bibinfo {year} {2020})},\
  \bibinfo {note} {[Erratum: Astron.Astrophys. 652, C4 (2021)]},\ \Eprint
  {https://arxiv.org/abs/1807.06209} {arXiv:1807.06209 [astro-ph.CO]}
  \BibitemShut {NoStop}%
\bibitem [{\citenamefont {Gianotti}\ \emph {et~al.}(2005)\citenamefont
  {Gianotti} \emph {et~al.}}]{Gianotti:2002xx}%
  \BibitemOpen
  \bibfield  {author} {\bibinfo {author} {\bibfnamefont {F.}~\bibnamefont
  {Gianotti}} \emph {et~al.},\ }\bibfield  {title} {\bibinfo {title} {{Physics
  potential and experimental challenges of the LHC luminosity upgrade}},\
  }\href {https://doi.org/10.1140/epjc/s2004-02061-6} {\bibfield  {journal}
  {\bibinfo  {journal} {Eur. Phys. J. C}\ }\textbf {\bibinfo {volume} {39}},\
  \bibinfo {pages} {293} (\bibinfo {year} {2005})},\ \Eprint
  {https://arxiv.org/abs/hep-ph/0204087} {arXiv:hep-ph/0204087} \BibitemShut
  {NoStop}%
\end{thebibliography}%
\end{document}